\documentclass[a4paper]{article}
\usepackage[affil-it]{authblk}
\usepackage{graphicx}
\usepackage{epstopdf}
\usepackage{times}
\usepackage{amsmath}
\usepackage{amssymb}
\usepackage{multirow}
\usepackage{bm}
\usepackage{todonotes}
\usepackage{amsfonts}
\usepackage{algorithmic}
\usepackage{textcomp}
\usepackage{xcolor}
\usepackage{ctable}
\usepackage{soul}
\usepackage[T1]{fontenc}
\usepackage{caption}
\usepackage{subcaption}
\usepackage{orcidlink}

\begin{document}
\title{Improving Autoencoder Training Performance for Hyperspectral Unmixing with Network Reinitialisation}

\author{Kamil Książek$^{1,2}$\orcidlink{0000-0002-0201-6220}, Przemysław Głomb$^{1}$\orcidlink{0000-0002-0215-4674}, Michał Romaszewski$^{1}$\orcidlink{0000-0002-8227-929X}, \\ Michał Cholewa$^{1}$\orcidlink{0000-0001-6549-1590}, Bartosz Grabowski$^{1}$\orcidlink{0000-0002-2364-6547}, Kriszti\'an B\'uza$^{3}$\orcidlink{0000-0002-7111-6452}
}
\affil{{\normalsize $^{1}$Institute of Theoretical and Applied Informatics, Polish Academy of Sciences, \\ 44-100 Gliwice, Poland \\
	\texttt{\{kksiazek, przemg, mromaszewski, mcholewa, \\ bgrabowski\}@iitis.pl}} \and
	{\normalsize $^{2}$Department of Data Sciences and Engineering, Silesian University of Technology, \\ 44-100 Gliwice, Poland \\
	\texttt{kamil.ksiazek@polsl.pl}} \and
	{\normalsize $^{3}$Biointelligence Group, Department of Mathematics-Informatics, \\ Sapientia Hungarian University of Transylvania, 540485 T\^argu Mureș, Romania \\
	\texttt{buza@biointelligence.hu}}}
\maketitle
\begin{abstract}
	Neural networks, in particular autoencoders, are one of the most \linebreak promising solutions for unmixing hyperspectral data, i.e. reconstructing the spectra of observed substances (endmembers) and their relative mixing fractions (abundances), which is needed for effective hyperspectral analysis and classification. However, as we show in this paper, the training of autoencoders for unmixing is highly dependent on weights initialisation; some sets of weights lead to degenerate or low-performance solutions, introducing negative bias in the expected performance. In this work, we experimentally investigate autoencoders stability as well as network reinitialisation methods based on coefficients of neurons' dead activations. We demonstrate that the proposed techniques have a positive effect on autoencoder training in terms of reconstruction, abundances and endmembers errors.
	%We demonstrate that proposed techniques have a positive effect on training metrics, both on reconstruction error and abundances or endmembers error.
	% \keywords{Autoencoders \and Hyperspectral unmixing \and Training stability \and Network reinitialisation.}
\end{abstract}
\section{Introduction}

Hyperspectral imaging (HSI) combines reflectance spectroscopy with image processing -- image pixels contain information about hundreds of spectral bands that can characterise chemical composition and properties of visible objects. In HSI the spectra of pixels are often a mixture of different substances~\cite{keshava2002unmixing}, as the sensor captures light reflected from nearby objects or aggregated from several sources due to the low spatial resolution. The task of hyperspectral unmixing (HU) is to reconstruct the original spectra of observed substances, called endmembers, and their fractional mixture coefficients, called abundances. On the one hand, HU facilitates further data analysis and improves classification results~\cite{Guo2021unmixing}. On the other hand, correlations between pixels and huge data volume resulting from the fact that every pixel can be treated as an example in a high-dimensional feature space, make neural networks particularly suitable models for HU. 

Although a number of machine learning algorithms for HU based on statistical and geometric principles have been developed~\cite{Bioucas2012overview}, 
for the aforementioned reasons, deep learning models seem to be the most promising solution, with autoencoders (AE) emerging as an architecture of choice. Most prominent examples include: a deep AE network~\cite{su2019DAEN} which is a sequence of stacked AE followed by a variational AE, generative and encoder models trained using pure pixel information~\cite{Borsoi2020generative}, EndNet architecture~\cite{ozkan2019endnet} or deep convolutional autoencoders~\cite{Palsson2020convolutional}.

One of key elements of training a neural model is a proper initialisation of weights, which prevents the phenomenon of vanishing or exploding gradients. Several weight initialisation methods have been introduced, e.g.~\cite{glorot2010xavier,He2015rectifiers}. However, while those methods are derived from examining gradient flow principles, they are usually applied without verification of the quality of initial weights. 
% As a result, when training fails there is no way of knowing if the failure results from a combination of training data and hyperparameter choice, or a bad initialisation. 

In this work we study the problem of AE training failures, resulting from bad initialisation weights, focusing on the problem of HU. We also present network reinitialisation methods which can alleviate a problem of bad weights and improve the network performance. In particular, we present the following contributions:
\begin{enumerate}
	\item We have experimentally verified the presence of failed trainings of autoencoders in HU scenarios. We have investigated this effect through $n=100000$ individual autoencoder training sessions across a diverse range of variables, and found that this effect persists across all studied variants of autoencoder architectures, datasets, weight initialisation methods, loss function types, and hyperparameter choices.
	\item To the best of our knowledge, this is the first detailed study of such failures in a standard autoencoder training scenario on a real-world hyperspectral dataset. 
	% In these cases, standard evaluation of hyperparameters would contaminate the score of hyperparameter values with bias resulting from unknown bad initialisations.
	\item Based on our results, we use the statistical analysis with the Kruskal-Wallis H-test to empirically confirm the thesis that a specific autoencoder initialisation affects the final data reconstruction error of a trained model.
	\item To resolve this issue, we propose network reinitialisation methods based on dead activations' coefficients. We show that for networks with ReLU activation function these approaches can both mitigate the impact of bad weights initialisations as well as unfavourable weight values resulting from the training.
\end{enumerate}

\subsection{Related work}
An overview of the HU methods can be found in ~\cite{Bioucas2012overview}. The approaches range from simple pure-pixel algorithms e.g. Pixel Purity Index (PPI)~\cite{Boardman1995PPI} or  N-FINDR~\cite{Winter1999NFINDR}
% or VCA~\cite{2005NascimentoVCA}. 
to more complex ones e.g. SISAL~\cite{Bioucas2009SISAL} which work with non-pure pixels and noisy data.% The SISAL algorithm~\cite{Bioucas2009SISAL} tries to solve it through a nonconvex optimisation, where the hard nonnegativity constraint for abundances is replaced by a regulariser which makes the method more robust.

An autoencoder (AE) is a neural network that through hidden layers compresses an input into a lower-dimensional (latent) space and reconstructs the original data. Reduction of the input dimensionality makes the AE well-suited for HU, thus they are often used as a base for HU algorithms. In~\cite{palsson2018unmixing}, authors analysed fully connected AE-based architectures for blind unmixing in an unsupervised setting. The use of AE for unmixing in a nonlinear case was a focus of~\cite{zhao2019hyperspectral}, where authors showed how in certain situation the linearity assumption will not hold. The approach via convolutional AE was tested in~\cite{palsson2019convoae} and~\cite{ranasinghe2020convolutional} however, both approaches use spectral and spatial features of data at the same time.

The general problem of random weight initialisation leading to inadequate results was observed in~\cite{hinton2006reducing}, \cite{krizhevsky2011RBM} and mitigated by the use of a stacked Restricted Boltzmann Machines (RBMs) to determine the initial weights for AE networks. In~\cite{glorot2010xavier}, a weight initialisation scheme was proposed that maintains activation and back-propagated gradients variance as one moves up or down the network. % Authors of~\cite{He2015rectifiers}, to alleviate the vanishing gradients when method  from~\cite{glorot2010xavier} is used for ReLU (non-linear) function, proposed a method based on zero-mean Gaussian distribution with $\sigma$ dependent on the number of pixels and input channels of a given layer. In addition, Parametric Rectified Linear Unit (PReLU) activation function with adaptive coefficients on the negative part of the real axis was tested. In~\cite{Yun2020begin} authors investigated influence of random initialisation and methods from ~\cite{glorot2010xavier} and~\cite{He2015rectifiers} on DNN learning. Their analysis yields some insights into learning dynamics of DNNs, but it does not indicate a clear winning method for weight initialisation.

Authors of~\cite{lecun2000backprop} discuss convergence of back-propagation, using several heuristics to support NN construction.  In~\cite{Plaza2006unmixing} authors propose to mitigate the instability of the unmixing algorithm with multiple initialisations. In~\cite{su2019DAEN} a scheme is proposed, where the first layer of a neural network determines the initialisation parameters for the unmixing engine, similarly to~\cite{guo2015cascade} where a cascade model is proposed. In~\cite{ozkan2019endnet} the EndNet algorithm is paired with VCA (Vertex Component Analysis) filter or FCLS (Fully Constrained Least Squares) for initialisation. The FCLS is also used in~\cite{Borsoi2020generative}. Results of the first run of the network may be used to re-initialise it in order to improve results~\cite{Lv2017facial}. 

We point out that naive initialisation of the weights may lead to dead neurons. The problem of dead neurons is that certain neurons output $0$ regardless of the input. This makes them impossible to train using gradient-based optimisation methods. In~\cite{Lu2020neurons}, authors provide theoretical analysis regarding death of neurons with ReLU activation function. They prove that ReLU network will eventually die as its depth goes to infinity. To alleviate the problem of dying neurons, they propose a new weight initialization procedure. In~\cite{Rister2021probabilistic}, authors show that it is possible to increase the network depth with guaranteed probability of living weights initialisation, as long as the network width increases accordingly. They also propose a sign flipping scheme to make sure the ratio of living data points in a $k$-layer network is at least $2^{-k}$.

Various attempts are made to improve the optimization of weights. In~\cite{Alabdulmohsin2021impact}, authors study the impact of different reinitialisation methods on generalisation using multiple convolutional neural networks (CNNs). In~\cite{Bingham2021autoinit} the AutoInit, an algorithm for dynamic scaling of weights of neural networks is proposed. Its potential is demonstrated on various architectures, including CNNs or residual networks.

%, as well as 12 image classification datasets. They also introduce their own layerwise reinitialisation algorithm.

\section{Performance investigation of autoencoders in a hyperspectral unmixing problem}

\subsection{Linear spectral mixing}
In this work we use the  Linear Mixing Model (LMM) of the pixel spectra, i.e. a $B$-band pixel $\bm{x} = \begin{bmatrix}x_{1}, ..., x_{B}\end{bmatrix}^\top$ is written as a linear combination of $E$ endmembers with the addition of a noise vector, i.e. $\bm{x} = \sum\limits_{j=1}^{E} a_{j} \cdot \bm{w_{j}} + \bm{n}$, where $\bm{a} = \begin{bmatrix}a_{1}, ..., a_{E}\end{bmatrix}^\top$ is a vector of abundances, $\boldsymbol{W} = \begin{bmatrix}\bm{w_{1}}, ..., \bm{w_{E}}\end{bmatrix}$ is a matrix of endmembers, $\boldsymbol{W} \in \mathbb{R}^{B \times E}$ and $\bm{n} = \begin{bmatrix}n_{1}, ..., n_{B}\end{bmatrix}^\top$ is a noise vector. To preserve the physical properties of abundances, it is necessary to ensure that the nonnegativity constraint is fulfilled and the sum of all fractional abundances equals to one. It means that $\forall j \in \lbrace 1, ..., E \rbrace ~~ a_{j} \geq 0,	\sum\limits_{j=1}^{E} a_{j} = 1$.

For $M$ pixels in the image $\boldsymbol{X} \in \mathbb{R}^{B \times M}$ the corresponding abundance matrix is denoted as $\boldsymbol{A} = \begin{bmatrix}\bm{a_{1}}, ..., \bm{a_{M}}\end{bmatrix}$, $\boldsymbol{A} \in \mathbb{R}^{E \times M}$ and $\boldsymbol{N} = \begin{bmatrix}\bm{n_{1}}, ..., \bm{n_{M}}\end{bmatrix}$, $\boldsymbol{N} \in \mathbb{R}^{B \times M}$ is the noise matrix, where $\bm{n_{i}} \in \mathbb{R}^{B \times 1}$ is a noise vector, $i \in \lbrace 1, ..., M \rbrace$. Accordingly, the LMM Equation can be rewritten as $\boldsymbol{X} = \boldsymbol{W}\boldsymbol{A} + \boldsymbol{N}$. The aim of the HU process is to estimate the endmembers matrix $\boldsymbol{W}$ and the abundances matrix $\boldsymbol{A}$ which provide an estimate of the pure spectra of substances and their fractions present in different pixels of the image.

\subsection{Architectures of autoencoders for hyperspectral unmixing}

\begin{figure*}[!t]
	\centering
	\includegraphics[trim={0,0cm, 0,2cm, 0,0cm, 0,2cm},clip,width=\textwidth]{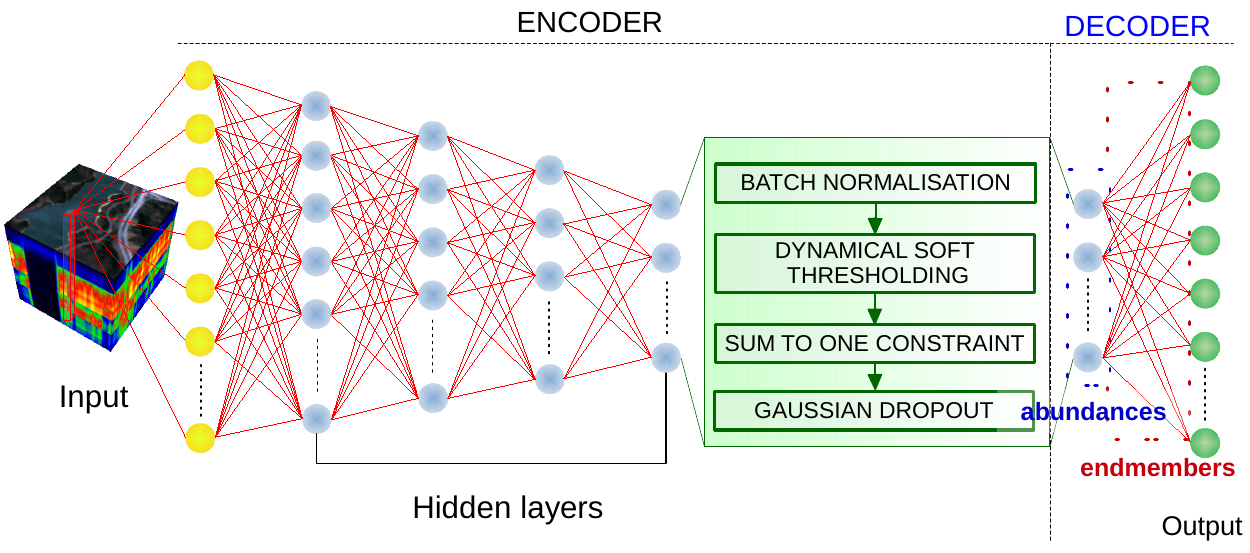}
	\caption{The pipeline of the autoencoder architecture from~\cite{palsson2018unmixing}, denoted as \emph{original}.}
	\label{fig:architecture}
\end{figure*}

We focus on the architecture presented in~\cite{palsson2018unmixing} (see Figure~\ref{fig:architecture}). Its encoder part consists of multiple linear layers that transform the input data into a latent space, according to the pattern:
%\begin{equation}
$
	\textrm{Enc}(\boldsymbol{X_{b}}): \mathbb{R}^{B \times b_{s}} \rightarrow \mathbb{R}^{E \times b_{s}},
$
%\end{equation}
where $\boldsymbol{X_{b}} \in \mathbb{R}^{B \times b_{s}}$ is a batch of input data, $b_{s}$ is a batch size. Then, in the decoder part, input spectra are reconstructed based on a latent representation:
%\begin{equation}
$
	\textrm{Dec}(\textrm{Enc}(\boldsymbol{X_{b}})): \mathbb{R}^{E \times b_{s}} \rightarrow \mathbb{R}^{B \times b_{s}}.
$
%\end{equation}

We denote the reconstructed image by $\boldsymbol{\hat{X}}~=~\textrm{Dec}(\textrm{Enc}(\boldsymbol{X}))$. The goal of the autoencoder is to minimise difference between $\boldsymbol{X}$ and $\boldsymbol{\hat{X}}$. The design of this architecture allows it to be used for HU; as a decoder has only one layer, its neurons' activations on the last encoder layer can be treated as abundance vectors while weights connecting the encoder part with the output of the autoencoder can be considered as endmembers. 

The authors of~\cite{palsson2018unmixing} studied several architectures with different number of layers and activation functions. We focus on the one that achieved one of the highest efficiency, i.e. a version with sigmoid activation function. This architecture consists of four linear layers in the encoder part, having $9E$, $6E$, $3E$ and $E$ neurons, respectively. After that, a batch normalisation (BN) layer is applied. Then, a dynamical soft thresholding (ST) is used which can be written as
%\begin{equation}
$
	\boldsymbol{x_{ST}^{i}} = \max(\boldsymbol{0}, \boldsymbol{x_{BN}^{i}} - \boldsymbol{\alpha}),
$
%\end{equation}
where $i \in \lbrace 1, ..., b_{s} \rbrace$, $\boldsymbol{X_{ST}} = [\boldsymbol{x_{ST}^{1}}, ..., \boldsymbol{x_{ST}^{b_{s}}}]$, $\boldsymbol{X_{BN}} = [\boldsymbol{x_{BN}^{1}}, ..., \boldsymbol{x_{BN}^{b_{s}}}]$ are matrices with all batch pixels after soft thresholding or batch normalisation, respectively; $\boldsymbol{0}$ is a zero vector and $\boldsymbol{\alpha}$ is a vector of trainable parameters. Then, to ensure that the sum to one constraint for abundances is met, each pixel vector is normalised i.e.
%\begin{equation}
$
	\forall i \in \lbrace 1, ..., b_{s} \rbrace ~ \boldsymbol{x_{norm}^{i}} = {\boldsymbol{x_{ST}^{i}}}/{\left(\sum\limits_{k=1}^{E} x_{ST}^{i, k}\right)},
$
%\end{equation}
where $\boldsymbol{x_{norm}^{i}}$ is the $i$--th batch vector after normalisation, $x_{ST}^{i, k}$ is the $k$--th coordinate of the $i$--th vector. Gaussian Dropout (GD) is applied as a last encoder layer, but only during training. Finally, the single decoder layer reconstructs the signal from the latent space to the input space. We denote this architecture as the \emph{original}.

To investigate the underlying dependencies of AEs for HU we have also prepared a simplified version of the described architecture, denoted \emph{basic}, where an encoder has two linear layers in which ReLU activation function is used. The number of neurons in the first hidden layer is a hyperparameter and is equal to $n_{1} E$, where $E$ is the number of endmembers, while the second hidden layer ending the encoder part has $E$ neurons. There are no BN, ST or GD layers. A normalisation layer is left to ensure that the sum of fractional abundances per each pixel is equal to one. A decoder part has one linear layer.

\subsection{Performance evaluation}

\subsubsection{Datasets}

We used two well-known datasets for HU: Samson and Jasper Ridge~\cite{zhu2017hyperspectral}.

% \begin{figure}[!h]
% 	\centering
% 	\begin{subfigure}[b]{0.2\textwidth}
% 		\includegraphics[trim={1,75cm, 0,1cm, 1,75cm, 0,1cm},clip,width=1.0\textwidth]{graphics/samson.pdf}
% 		\caption{RGB image of the Samson dataset \label{fig:samson_rgb}}
% 	\end{subfigure}
% 	\begin{subfigure}[b]{0.25\textwidth}
% 		\includegraphics[trim={0,35cm, 0,55cm, 0,4cm, 0,4cm},clip,scale=0.41]{graphics/endmembers_GT_Samson.pdf}
% 		\caption{Spectra of Samson dataset endmembers \label{fig:samson_gt}}
% 	\end{subfigure}%
% 	\caption{The presentation of the Samson dataset which consists three endmembers: water, tree and soil.}
% 	\label{fig:samson}
% \end{figure}

Samson is an image with dimensions $95 \times 95 \times 156$, spectral range of $401-889~\mathrm{nm}$ and spectral resolution  $\sim3.13~\mathrm{nm}$. Pixels spectra are a mixture of three endmembers: water, trees and soil. %(see Figure~\ref{fig:samson}).

% \begin{figure}[!h]
% 	\centering
% 	\begin{subfigure}[b]{0.2\textwidth}
% 		\includegraphics[trim={1,75cm, 0,1cm, 1,75cm, 0,1cm},clip,width=1.0\textwidth]{graphics/jasper_ridge.pdf}
% 		\caption{RGB image of the Jasper Ridge dataset \label{fig:jasper_rgb}}
% 	\end{subfigure}
% 	\begin{subfigure}[b]{0.25\textwidth}
% 		\includegraphics[trim={0,35cm, 0,55cm, 0,4cm, 0,4cm},clip,scale=0.41]{graphics/endmembers_GT_Jasper.pdf}
% 		\caption{Spectra of Jasper Ridge dataset endmembers \label{fig:jasper_gt}}
% 	\end{subfigure}%
% 	\caption{The presentation of the Jasper Ridge dataset which consists four endmembers: water, tree, soil and road.}
% 	\label{fig:jasper}
% \end{figure}

Jasper Ridge is an image with dimensions $100 \times 100 \times 198$. Originally, there were $224$ bands covering the range of  $380-2500~\mathrm{nm}$ but bands $1-3$, $108-112$, $154-166$ and $220-224$ were removed due to disturbances, as did authors in~\cite{ozkan2019endnet,zhu2017hyperspectral}. Endmembers represent: trees, water, soil and road. % (see Figure~\ref{fig:jasper}).  

\subsubsection{Experiment}

The goal of the experiment is to confirm the impact of weight initialisation on the final reconstruction error of the AE.
For each of $N=50$ randomly initialised AEs, we have performed $k=50$ separate training sessions for each AE initialisation with the same dataset and hyperparameter values, resulting in $n_a=2500$ trained models. This has been repeated across ten different hyperparameters sets (see Table~\ref{tab:parameters}) and four methods of weight initialisation, i.e. He~\cite{He2015rectifiers} and Glorot~\cite{glorot2010xavier} with normal or uniform distribution, bringing the total number of models to $n_b = 100 000$.

Each model is evaluated as follows. Let $D$ be a HSI image and let $V_{GT}$ be a set of correct endmembers for image $D$. Let also $D_{GT}$ be ground truth for fractional abundances, i.e. correct abundances for endmembers $V_{GT}$. The AE $A_{i,j}, i = 1, \ldots, N, j= 1, \ldots, k$ is trained with a set of vectors from dataset $D$. The endmembers $V_{i,j}$ and abundance vectors are then extracted; the endmembers set $V_{i,j}$ is matched to $V_{GT}$. The match used is the permutation of endmembers with smallest distance to $V_{GT}$.  Abundance vectors are compared to ground truth image $D_{GT}$ to calculate error $E^a_{i,j}$ in terms of RMSE, like in~\cite{su2019DAEN}, while endmembers are compared to their counterparts from $V_{GT}$ and the value of $E^e_{i,j}$ is calculated according to the SAD function. For a given experiment mean endmembers error is calculated, as follows: $E^{e} = \sum\nolimits_{i=1}^{N} \sum\nolimits_{j=1}^{k} \frac{E^{e}_{i,j}}{Nk}$. Mean abundances error $E^{a}$ is calculated analogously.

\subsection{Parameters}

We have explored the effect of weight initialisation with a range of hyperparameters (see Table~\ref{tab:parameters}). We have used: two architectures (\emph{original} and  \emph{basic}), two datasets, two loss functions (MSE and SAD). The rest of hyperparameters (e.g. learning rate, batch size, GD parameter etc.) have been tuned using RayTune optimisation library~\cite{liaw2018tune}. For the comparison, we also included hyperparameters indicated by the authors of the~\emph{original} architecture~\cite{palsson2018unmixing}. In all experiments we have used the Adam optimiser. Additionally, we have used four initialisation algorithms from~\cite{He2015rectifiers} and~\cite{glorot2010xavier}, each with uniform\footnote{With \cite{He2015rectifiers} method using uniform initialisation we have used the version from the PyTorch library, which differs from the paper with: 1) biases are not initialised to 0; 2) bounds of a uniform distribution are constant and not dependent on the number of connections.} and normal distribution. The source code necessary for replication of experiments is publicly available in our github repository\footnote{The source code is available at the following link: \url{https://github.com/iitis/AutoencoderTestingEnvironment}.}.

%When using autoencoders, there is a number of hyperparameters to be decided. As they have an influence on the result, their choice is significant for general effectiveness of the algorithm. However we argue, that the what is crucial for our experiment is set them on any sensible level, not necessarily optimal. The main importance is for various initialisation work on a same set of hyperparameters. To verify our hypothesis, we performed 10 different experiment scenarios, using two described architectures: the \emph{original} and the \emph{basic}, as well as two loss functions: MSE and SAD. We selected two datasets for unmixing: Samson and Jasper Ridge. The hyperparameters sets were chosen using RayTune optimisation library \cite{liaw2018tune}. For the comparison purpose, we also added hyperparameters indicated by the authors of the \emph{original} architecture \cite{palsson2018unmixing}. In all experiments we used Adam optimiser. 

\setlength{\tabcolsep}{3pt}
\ctable[
	caption = {The list of performed experiments with all hyperparameters used. An encoder means the number of neurons on the first hidden layer and it concerns only the \emph{basic} architecture. Gaussian Dropout (GD) is applied only for the \emph{original} architecture.},
	label = tab:parameters,
	mincapwidth = \textwidth,
	doinside = {\scriptsize},
	star = True,
	pos = t
]{cccccccll}{}{\FL
	% \shortstack{a1 \\ a2}
	experiment ID & architecture & hyperparameters origin & loss & dataset & encoder & batch size & learning rate & GD \ML
	1 & \emph{original} & RayTune & MSE & Samson & $-$ & 100 & 0.01 & 0. \\
	2 & \emph{original} & RayTune & SAD & Samson & $-$ & 100 & 0.01 & 0. \\
	3 & \emph{original} & article & SAD & Samson & $-$ & 20 & 0.01 & 0.1 \\
	4 & \emph{basic} & RayTune & MSE & Samson & $10E$ & 4 & 0.0001 & $-$ \\
	5 & \emph{basic} & RayTune & SAD & Samson & $20E$ & 4 & 0.0001 & $-$ \\
	6 & \emph{original} & RayTune & MSE & Jasper Ridge & $-$ & 100 & 0.01 & 0. \\
	7 & \emph{original} & RayTune & SAD & Jasper Ridge & $-$ & 100 & 0.01 & 0. \\
	8 & \emph{original} & article & MSE & Jasper Ridge & $-$ & 5 & 0.01 & 0.1 \\
	9 & \emph{original} & article & SAD & Jasper Ridge & $-$ & 5 & 0.01 & 0.1 \\
	10 & \emph{basic} & RayTune & MSE & Jasper Ridge & $10E$ & 20 & 0.001 & $-$ \LL
	% 11 & \emph{basic} & RayTune & SAD & Jasper Ridge & $10E$ & 4 & 0.0001 & 1e-05 & $-$ \LL
}

\subsection{Statistical verification}
To confirm a non-uniform behaviour of weights (the existence of initialisations leading to worse models after the network training than other initialisations) we have used the Kruskal-Wallis H-test for a one-way analysis of variance~\cite{kruskall1952test}. The test is performed as follows: our models are treated as $N$ different populations where every population corresponds to a single set of initial weights and samples  correspond to error estimates of consecutive training runs. The hypothesis of a H-test are as follows:
%It is an extension of Mann-Whitney test for two independent samples~\cite{washington2001statistics}. 
% We have not used the ANOVA parametric test since the assumption of variance equality among the populations was not met according to results of the Levene's test~\cite{levene1961robust}.

\emph{$H_{0}$}: All population means are equal, i.e. $\mu_{A_{1}} = \mu_{A_{2}} = ... = \mu_{A_{N}}$.

\emph{$H_{1}$}: At least one population has a statistically significantly different mean than the others.

%\noindent The value $\mu_{A_{i}},$ where $i~\in~\lbrace 1, ..., N \rbrace$, is the mean of the $i$--th population, i.e. mean results for $i$--th weights initialisation of the autoencoder. Generally, the test statistic is calculated based on the following equation~\cite{kruskall1952test}:
%\begin{equation}
%H = \dfrac{12}{C(C+1)} %\sum\limits_{j=1}^{N} %\dfrac{R_{j}^{2}}{n_{j}} - 3(C+1),
%\label{KruskalWallis_1}
%\end{equation}
%where $R_{j}$ is the sum of ranks of the $j$--th population, $j~\in~\lbrace 1, ..., N \rbrace$. In our case $C = k \cdot N$ and $n_{1} = ... = n_{N} = k$. It is also recommended to correct the general formula of the H-test if some observations occur multiple times \cite{kruskall1952test, siegel1956statistics}. In that case, mean of the ranks for all tied observations is derived and a correction of the above formula is prepared as in~\cite{kruskall1952test}.

%For a sufficiently large number of samples, if the null hypothesis is true, H-statistic has $\chi^2$ distribution with $N - 1$ degrees of freedom. The critical value can be read from the tables of $\chi^2$ distribution.

Since rejection of the null hypothesis does not give an answer which population differs from  others, a post-hoc analysis is performed using the Conover-Iman test \cite{conover1998practical,conover1979multiple} which can be used if and only if the null hypothesis of the Kruskal-Wallis H-test is rejected.
%Two populations $i$ and $j$, where $i, j \in \lbrace 1, ..., N \rbrace$, are different when the following inequality is true~\cite{washington2001statistics}:
%\begin{equation}
%\left\lvert \dfrac{R_{i}}{n_{i}} - \dfrac{R_{j}}{n_{j}} \right\rvert > t_{1-\frac{\alpha}{2}} \sqrt{S^{2} \cdot \dfrac{C - 1 - H}{C - N} \cdot \left( \dfrac{1}{n_{i}} + \dfrac{1}{n_{j}} \right)}, 
%\label{Conover_test}\end{equation}
%where $H$ is value of the statistic test from Equation \ref{KruskalWallis_1}, $t_{1-\frac{\alpha}{2}}$ is a value of $(1 - \frac{\alpha}{2})$ quantile of $t$ distribution, $\alpha$ is the significance level and $n_{i}, n_{j}$ is the number of samples in the $i$--th and $j$--th population, respectively. In our case each population has $k$ samples. $S^{2}$ can be expressed in the following way~\cite{washington2001statistics}:
%\begin{equation}
%S^{2} = \dfrac{1}{C - 1} \left( \sum\limits_{i=1}^{N} \sum\limits_{j=1}^{n_{i}} R_{i,j}^{2} - \dfrac{C \cdot (C + 1)^2}{4} \right),
%\end{equation}
%where $R_{i,j}$ is a rank value of $j$--th sample of the $i$--th population. 
By performing the pairwise comparison for all population pairs, we can conclude which differences between populations are statistically significant.

\subsection{Network reinitialisation method}
\label{sec:reinit}
To minimise the impact of bad neural network weights we propose statistics that allow us to detect and alleviate this phenomenon. % It should avoid wasting computation time and improve the network performance.

% \subsubsection{Definitions}

Let $\mathfrak{N}$ be a $n-$layer autoencoder network for which $\mathbf{c} = [c_{0}, c_{1}, c_{2}, ..., c_{n}]$ determines the number of neurons on subsequent layers. A number $c_{0}$ denotes the input size while for $i \in \lbrace 1, 2, ..., n \rbrace$, $c_{i}$ is the number of neurons in the $i$--th network layer. We assume that ReLU activation function is used in all hidden layers. Let us also define a matrix of neurons activation values of the $i$--th layer of $\mathfrak{N}$, $\mathbf{G}_{i}(\boldsymbol{X_{b}}) = [\mathbf{g_{1}}, ..., \mathbf{g_{b_{s}}}]$, where $\boldsymbol{X_{b}} \in \mathbb{R}^{B \times b_{s}}$ is a batch of network input data and $i \in \lbrace 1, 2, ..., n \rbrace$. Columns of the matrix $\mathbf{G}_{i}(\boldsymbol{X})$ store vectors of neurons' activations values for consecutive input data points, i.e. $\mathbf{g_{j}} = [g_{j, 1}, g_{j, 2}, ..., g_{j, c_{i}}]^\top$, where $j \in \lbrace 1, ..., b_{s} \rbrace$.

The $q$--th neuron of the $k$--th layer is called \emph{dead for a given input data} $\mathbf{x_{i}}$, $\mathbf{x_{i}} \neq \mathbf{0}$, where $\mathbf{x_{i}}$ is an element of $\boldsymbol{X_{b}}$, if $g_{j, q} = 0$ for $\mathbf{x_{i}}$ as the input of the network, $q \in \lbrace 1, ..., c_{k} \rbrace$ and $k \in \lbrace 1, ..., n \rbrace$. During a given training iteration we have $c_{k} \cdot b_{s}$ activations values for the $k$--th network layer. We introduce a dead activations' coefficient for the $k$--th layer, $d_{dead}^{k}$:
\begin{equation}
	d_{dead}^{k} = \frac{\mathcal{N}_{0}^{k}}{c_{k} \cdot b_{s}} \in [0, 1],
\end{equation}
\noindent where $\mathcal{N}_{0}^{k}$ is a number of zero activations for all neurons of the $k$--th layer. We also calculate a dead activations' coefficient for the $q$--th neuron of the $k$--th network layer, $d_{dead}^{k, q}$:
\begin{equation}
	d_{dead}^{k, q} = \frac{\mathcal{N}_{0}^{k, q}}{b_{s}} \in [0, 1],
\end{equation}
\noindent where $\mathcal{N}_{0}^{k, q}$ is a number of zero activations for the $q$--th neuron of the $k$--th network layer. During preliminary research using \emph{basic} architecture, it was found that there is a significant correlation between the number of dead activations for the second encoder layer in a selected run of the trained model and mean reconstruction error. In such AE architectures for HU the last encoder layer has only few neurons, according to the number of endmembers in a given dataset (e.g. 3 for Samson and 4 for Jasper Ridge) so this layer is a bottleneck. We investigated models generated during Experiments 4 and 10 (according to Table~\ref{tab:parameters}), i.e. using MSE loss function and two HSI datasets. Results of Spearmank's rank correlation coefficient for all weight initialisation methods ranged between $0.76$ and $0.89$. Based on above observations we have proposed three network reinitialisation methods dependent on the number of dead activations for the second encoder layer ($d_{dead}^{2}$) or the number of dead activations for consecutive neurons of the second encoder layer, i.e. $( d_{dead}^{2, 1}, ..., d_{dead}^{2, c_{2}})$:

\begin{itemize}
	\item \emph{whole network reinitialisation}: if $d_{dead}^{2} > t$ then all model's weights are randomly generated according to the given weights initialisation approach;
	\item \emph{single layer reinitialisation}: if $d_{dead}^{2} > t$ then all weights of the second encoder layer are reinitialised;
	\item \emph{partial reinitialisation of a single layer}: weights of the second encoder layer connected with neurons which exceeded the dead activations' thresholds are reinitialised.
\end{itemize}

\noindent During each iteration of the network training one of the above conditions is verified, depending on the selected method.

\section{Results}

\setlength{\tabcolsep}{2,4pt}
\ctable[
	caption = {Results for different weight initialisation methods: He~\cite{He2015rectifiers} and Glorot~\cite{glorot2010xavier} with normal / uniform distribution. For each experiment a H-test statistic and a logarithm of p-value are presented. The significance level $\alpha$ is equal to $0.05$. The `ph' column corresponds to the ratio of $\text{p-values}<\alpha$ in post-hoc analysis. Bold font indicates experiment with $\text{p-value}>\alpha$. In the case of very small p-values, log p-values are denoted `$-\inf$'.},
	label = tab:results,
	doinside = {\scriptsize},
	mincapwidth = \textwidth,
	star = True,
	pos = t
]{crlcrlcrlcrlc}{}{\FL
	% \shortstack{a1 \\ a2}
	init. & \multicolumn{3}{c}{He normal (KHN)} & \multicolumn{3}{c}{He uniform (KHU)} & \multicolumn{3}{c}{Glorot normal (XGN)} & \multicolumn{3}{c}{Glorot uniform (XGU)} \ML
	exp. ID & H-stat & log p-val&ph& H-stat & log p-val&ph & H-stat & log p-val&ph & H-stat & log p-val&ph \ML
	1 & $200.9$ & $-45.07$ & $0.65$ & $243.2$ & $-61.76$ & $0.70$ & $78.8$ & $-5.41$ & $0.55$ & $70.4$ & $-3.72$ & $0.54$ \NN
	2 & $891.2$ & $-355.38$ & $0.84$ & $443.2$ & $-147.76$ & $0.77$ & $625.9$ & $-231.04$ & $0.81$ & $660.3$ & $-246.96$ & $0.82$ \NN
	3 & $763.8$ & $-295.33$ & $0.83$ & $267.8$ & $-71.82$ & $0.69$ & $239.1$ & $-60.11$ & $0.69$ & $180.9$ & $-37.49$ & $0.66$ \NN
	4 & $2185.8$ & $-\inf$ & $0.93$ & $2025.3$ & $-\inf$ & $0.85$ & $1997.6$ & $-\inf$ & $0.87$ & $2141.5$ & $-\inf$ & $0.87$ \NN
	5 & $1954.3$ & $-\inf$ & $0.93$ & $2093.4$ & $-\inf$ & $0.94$ & $1840.8$ & $-\inf$ & $0.92$ & $1777.2$ & $-\inf$ & $0.91$ \NN
	6 & $134.0$ & $-20.95$ & $0.61$ & $75.8$ & $-4.79$ & $0.55$ & $76.3$ & $-4.90$ & $0.55$ & $98.8$ & $-10.32$ & $0.57$ \NN
	7 & $903.8$ & $-361.36$ & $0.85$ & $761.1$ & $-294.07$ & $0.82$ & $953.8$ & $-385.10$ & $0.84$ & $871.0$ & $-345.85$ & $0.83$ \NN
	8 & $77.3$ & $-5.09$ & $0.55$ & $75.4$ & $-4.71$ & $0.54$ & $78.4$ & $-5.33$ & $0.55$ & $93.3$ & $-8.88$ & $0.57$ \NN
	9 & $69.2$ & $-3.50$ & $0.54$ & $\mathbf{66.2}$ & $\mathbf{-2.98}$ & $-$ & $74.2$ & $-4.46$ & $0.54$ & $\mathbf{47.9}$ & $\mathbf{-0.65}$ & $-$ \NN
	10 & $1767.3$ & $-\inf$ & $0.91$ & $2041.9$ & $-\inf$ & $0.90$ & $1344.6$ & $-572.45$ & $0.84$ & $1155.1$ & $-481.29$ & $0.85$ \LL
}

% \begin{figure*}[!h]
% 	\centering
% 	\begin{subfigure}[b]{0.23\textwidth}
% 		\includegraphics[trim={2,8cm, 0,6cm, 3,3cm, 1,2cm},clip,width=1.0\linewidth]{graphics/post_hoc/F004_KHN_04022021_post_hoc_test.pdf}
% 		\caption{He normal init.}
% 	\end{subfigure}%
% 	\begin{subfigure}[b]{0.23\textwidth}
% 		\includegraphics[trim={2,8cm, 0,6cm, 3,3cm, 1,2cm},clip,width=1.0\linewidth]{graphics/post_hoc/F004_KHU_04022021_post_hoc_test.pdf}
% 		\caption{He uniform init.}
% 	\end{subfigure}%
% 	\begin{subfigure}[b]{0.23\textwidth}
% 		\includegraphics[trim={2,8cm, 0,6cm, 3,3cm, 1,2cm},clip,width=1.0\linewidth]{graphics/post_hoc/F004_XGN_04022021_post_hoc_test.pdf}
% 		\caption{Glorot normal init.}
% 	\end{subfigure}%
% 	\begin{subfigure}[b]{0.31\textwidth}
% 		\includegraphics[trim={2,8cm, 0,6cm, 0cm, 1,2cm},clip,width=1.0\linewidth]{graphics/post_hoc/F004_XGU_04022021_post_hoc_test.pdf}
% 		\caption{Glorot uniform init.}
% 	\end{subfigure}
% 	\caption{Results of post-hoc Conover-Iman test for pairs of initialisations in the Experiment 4.}
% 	\label{fig:post_hoc_exp_4}
% \end{figure*}

Results of the experimental evaluation of non-uniform behaviour of weights are presented in Table~\ref{tab:results} as Kruskal-Wallis H-test statistics and corresponding p-values for the significance level $\alpha = 0.05$. In all but one experiment $H_{0}$ was rejected due to p-values lower than $0.05$. An alternative hypothesis $H_{1}$ states that at least one initialisation of initial weights resulted in the value of reconstruction error that was significantly different from the rest. This confirms that the reconstruction error of the trained network depends on weight initialisation. 

\begin{figure*}[t]
	\centering
	% \begin{subfigure}[b]{0.23\textwidth}
	% 	\includegraphics[trim={2,8cm, 0,6cm, 3,3cm, 1,2cm},clip,width=1.0\linewidth]{graphics/post_hoc/F001_KHN_04022021_post_hoc_test.pdf}
	% \end{subfigure}%
	\begin{subfigure}[b]{0.24\textwidth}
		\includegraphics[trim={2,8cm, 0,6cm, 3,3cm, 1,2cm},clip,width=1.0\linewidth]{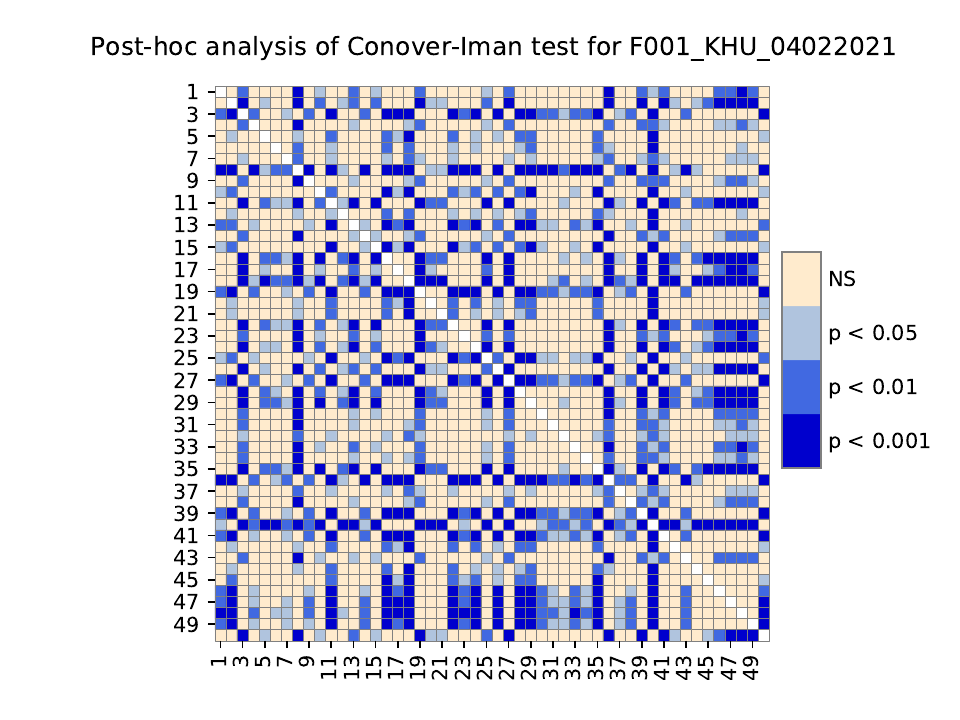}
		\caption{KHU, exp. no 1}
	\end{subfigure}%
	% \begin{subfigure}[b]{0.23\textwidth}
	% 	\includegraphics[trim={2,8cm, 0,6cm, 3,3cm, 1,2cm},clip,width=1.0\linewidth]{graphics/post_hoc/F001_XGN_04022021_post_hoc_test.pdf}
	% \end{subfigure}%
	\begin{subfigure}[b]{0.24\textwidth}
		\includegraphics[trim={2,8cm, 0,6cm, 3,3cm, 1,2cm},clip,width=1.0\linewidth]{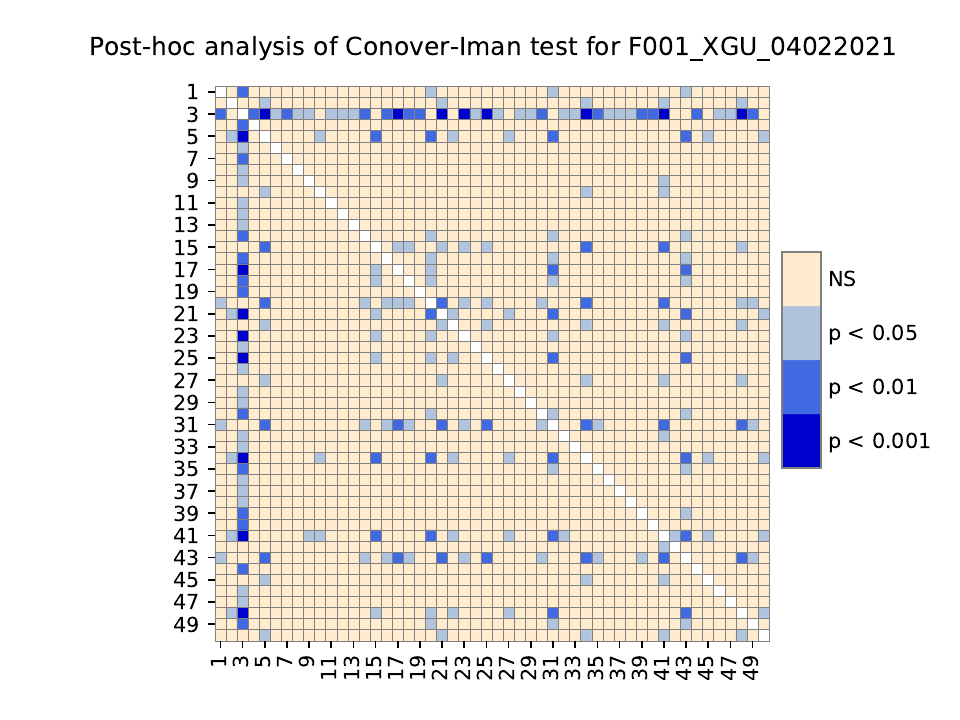}
		\caption{XGU, exp. no 1}
	\end{subfigure}%
	% \begin{subfigure}[b]{0.23\textwidth}
	% 	\includegraphics[trim={2,8cm, 0,6cm, 3,3cm, 1,2cm},clip,width=1.0\linewidth]{graphics/post_hoc/F004_KHN_04022021_post_hoc_test.pdf}
	% 	\caption{He normal init.}
	% \end{subfigure}%
	\begin{subfigure}[b]{0.24\textwidth}
		\includegraphics[trim={2,8cm, 0,6cm, 3,3cm, 1,2cm},clip,width=1.0\linewidth]{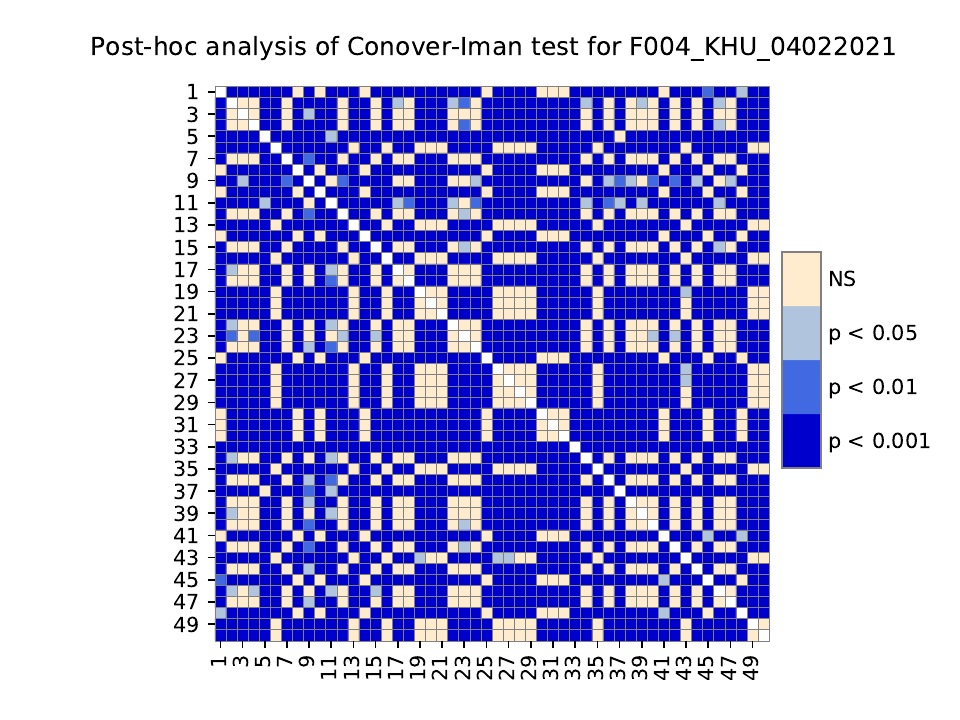}
		\caption{KHU, exp. no 4}
	\end{subfigure}%
	% \begin{subfigure}[b]{0.23\textwidth}
	% 	\includegraphics[trim={2,8cm, 0,6cm, 3,3cm, 1,2cm},clip,width=1.0\linewidth]{graphics/post_hoc/F004_XGN_04022021_post_hoc_test.pdf}
	% 	\caption{Glorot normal init.}
	% \end{subfigure}%
	\begin{subfigure}[b]{0.32\textwidth}
		\includegraphics[trim={2,8cm, 0,6cm, 0cm, 1,2cm},clip,width=1.0\linewidth]{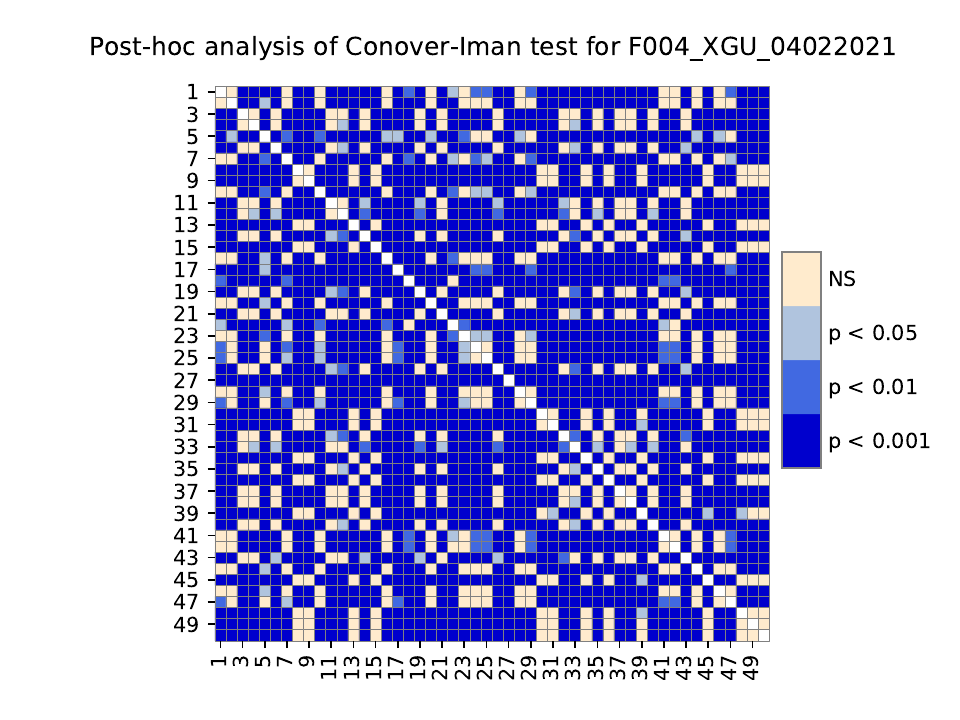}
		\caption{XGU, exp. no 4~~~~~~~~~~~~}
	\end{subfigure}
	\caption{Results of post-hoc Conover-Iman test for pairs of initialisations.}
	\label{fig:post_hoc_exp_1_4}
\end{figure*}

To compare individual initialisations, a post-hoc analysis with the Conover-Iman test was performed. Example results are presented in Figure~\ref{fig:post_hoc_exp_1_4} as heat-maps where each cell represents the statistical significance of difference between RMSE values of a pair of experiments. \emph{NS} denotes a statistically insignificant difference. An analysis of these matrices reveals that for some cases e.g. Experiment~1, there exists a subset of outlying initialisations while e.g. for Experiment~4 the majority of initialisation pairs are significantly different. A summary of post-hoc analysis is presented in Table~\ref{tab:results}. Values in `ph' columns are ratios of statistically significant ($p<0.05$) differences between individual experiments in accordance to the Conover-Iman test. Results indicate that using KHN initialisation slightly increased the number of outlying initialisations.

\subsection{Autoencoder training improvement}
We performed experiments validating the efficiency of network reinitialisation methods described in Section~\ref{sec:reinit}. We used 200 sets of initial weights from Experiment 4 (50 per each initialisation approach like XGN, XGU, KHN and KHU). Each model was trained once for each training improvement method. Table~\ref{tab:ReinitMethodsResults} presents mean results of described experiments with corresponding standard deviations. We denoted as \emph{baseline} initial training results, without the use of reinitialisation techniques. Furthermore, to check whether the improvement is statistically significant, the Wilcoxon signed-rank tests between results for baseline models and for models with the application of reinitialisation methods were performed. Conducted experiments clearly indicate that proposed reinitialisation techniques significantly outperform the baseline training approach. In the case of Glorot initialisations \emph{whole network} method achieved lowest RMSE and abundances error while for He algorithm \emph{partial reinitialisation} was on average the most efficient.
% The parametric t-test for pairs could not be used due to the fact that an assumption about data normality was not met, according to results of the Shapiro-Wilk test.  

\setlength{\tabcolsep}{6pt}
\ctable[
	caption = {Results of network reinitialisation methods. Consecutive rows present mean results with standard deviation for each network setup. Filled square ($\blacksquare$) means that the improvement between a given reintialisation method and baseline is statistically significant, in terms of Wilcoxon signed-rank test.},
	label = tab:ReinitMethodsResults,
	doinside = {\footnotesize},
	mincapwidth = \textwidth,
	star = False,
	pos = t
]{lllll}{}{\FL
	init. method & reinitialisation method & RMSE & abundances error & endmembers error \ML
	\multirow{5}{*}{XGN} & baseline & 0.036 $\pm$ 0.04 & 0.349 $\pm$ 0.04 & 0.744 $\pm$ 0.14 \\
	& partial reinitialisation & 0.089 $\pm$ 0.03 & 0.413 $\pm$ 0.08 & \textbf{0.451 $\pm$ 0.24} $\blacksquare$\\
	& single layer & 0.019 $\pm$ 0.03 $\blacksquare$ & 0.310 $\pm$ 0.06 $\blacksquare$ & 0.777 $\pm$ 0.19 \\
	& whole network & \textbf{0.007 $\pm$ 0.00} $\blacksquare$ & \textbf{0.276 $\pm$ 0.03} $\blacksquare$ & 0.687 $\pm$ 0.11 $\blacksquare$ \ML
	\multirow{5}{*}{XGU} & baseline & 0.052 $\pm$ 0.05 & 0.344 $\pm$ 0.05 & 0.723 $\pm$ 0.12 \\
	& partial reinitialisation & 0.089 $\pm$ 0.03 & 0.407 $\pm$ 0.08 & \textbf{0.375 $\pm$ 0.03} $\blacksquare$ \\
	& single layer & 0.020 $\pm$ 0.04 $\blacksquare$ & 0.307 $\pm$ 0.05 $\blacksquare$ & 0.707 $\pm$ 0.16 \\
	& whole network & \textbf{0.007 $\pm$ 0.00} $\blacksquare$ & \textbf{0.270 $\pm$ 0.03} $\blacksquare$ & 0.695 $\pm$ 0.11 \ML
	\multirow{5}{*}{KHN} & baseline & 0.047 $\pm$ 0.04 & 0.365 $\pm$ 0.03 & 1.176 $\pm$ 0.13 \\
	& partial reinitialisation & \textbf{0.024 $\pm$ 0.01} $\blacksquare$ & \textbf{0.330 $\pm$ 0.03} $\blacksquare$ & \textbf{1.100 $\pm$ 0.13} $\blacksquare$ \\
	& single layer & 0.028 $\pm$ 0.01 $\blacksquare$ & 0.355 $\pm$ 0.04 & 1.153 $\pm$ 0.11 \\
	& whole network & 0.027 $\pm$ 0.01 $\blacksquare$ & 0.355 $\pm$ 0.04 & 1.134 $\pm$ 0.13 $\blacksquare$ \ML
	\multirow{5}{*}{KHU} & baseline & 0.053 $\pm$ 0.05 & 0.357 $\pm$ 0.03 & 0.758 $\pm$ 0.19 \\
	& partial reinitialisation & \textbf{0.007 $\pm$ 0.00} $\blacksquare$ & \textbf{0.308 $\pm$ 0.04} $\blacksquare$ & 0.733 $\pm$ 0.12 \\
	& single layer & 0.014 $\pm$ 0.01 $\blacksquare$ & 0.347 $\pm$ 0.03 & 0.732 $\pm$ 0.15 \\
	& whole network & 0.008 $\pm$ 0.01 $\blacksquare$ & 0.316 $\pm$ 0.04 $\blacksquare$ & \textbf{0.663 $\pm$ 0.17} $\blacksquare$
	\LL}%

\subsection{Discussion}

We observed that optimisation using MSE as a loss function also minimises the reconstruction error in terms of SAD function, to some extent. This relationship is particularly evident in the case of \emph{basic} architecture. Indeed, if the value of MSE is close to 0, then also the SAD has to be close to 0. However, the opposite is not true, because the optimisation using SAD function does not reduce error in the sense of MSE. Moreover, in most cases, after training with SAD function, all or almost all reconstructed points are outside of the simplex designated by endmembers. This phenomenon occurs because SAD function is scale invariant which means that only the spectral angle between input and output points is minimised. It is not necessarily related to the reduction of Euclidean distance between spectra. Despite this, abundances error in terms of RMSE and endmembers SAD error are comparable or even lower than when training with the MSE function which can be seen in the detailed results in Table~\ref{tab:unmixing}.

\setlength{\tabcolsep}{2,4pt}
\ctable[
	caption = {Results of mean abundances error in terms of RMSE $(E^{a})$ and mean endmembers error in terms of SAD $(E^{e})$ with standard deviations for different weight initialisation methods: He~\cite{He2015rectifiers} and Glorot~\cite{glorot2010xavier} with normal / uniform distribution. Bold font indicates experiment with the lowest value of error.},
	label = tab:unmixing,
	doinside = {\scriptsize},
	mincapwidth = \textwidth,
	star = True,
	pos = t
]{ccccccccc}{}{\FL
	% \shortstack{a1 \\ a2}
	init. & \multicolumn{2}{c}{He normal (KHN)} & \multicolumn{2}{c}{He uniform (KHU)} & \multicolumn{2}{c}{Glorot normal (XGN)} & \multicolumn{2}{c}{Glorot uniform (XGU)} \ML
	exp. ID & abundances & endmembers & abundances & endmembers & abundances & endmembers & abundances & endmembers \ML
	1 & $0.34\pm0.0$ & $0.84\pm0.2$ & $0.36\pm0.0$ & $0.68\pm0.2$ & $\mathbf{0.32\pm0.0}$ & $\mathbf{0.36\pm0.3}$ & $0.33\pm0.1$ & $0.43\pm0.3$ \\
	2 & $0.19\pm0.1$ & $0.23\pm0.1$ & $0.13\pm0.1$ & $0.13\pm0.1$ & $0.10\pm0.1$ & $0.10\pm0.1$ & $\mathbf{0.10\pm0.1}$ & $\mathbf{0.10\pm0.1}$ \\
	3 & $\mathbf{0.06\pm0.1}$ & $0.05\pm0.1$ & $0.07\pm0.1$ & $0.04\pm0.1$ & $0.07\pm0.1$ & $0.05\pm0.1$ & $0.07\pm0.1$ & $\mathbf{0.04\pm0.1}$ \\
	4 & $0.36\pm0.0$ & $1.17\pm0.1$ & $0.36\pm0.0$ & $0.76\pm0.2$ & $0.35\pm0.0$ & $0.75\pm0.1$ & $\mathbf{0.34\pm0.0}$ & $\mathbf{0.73\pm0.1}$ \\
	5 & $0.35\pm0.0$ & $0.95\pm0.1$ & $0.35\pm0.0$ & $0.97\pm0.2$ & $\mathbf{0.34\pm0.0}$ & $0.88\pm0.2$ & $0.34\pm0.0$ & $\mathbf{0.87\pm0.1}$ \\
	6 & $0.29\pm0.0$ & $0.80\pm0.1$ & $0.24\pm0.1$ & $0.58\pm0.2$ & $0.22\pm0.0$ & $0.51\pm0.2$ & $\mathbf{0.22\pm0.0}$ & $\mathbf{0.50\pm0.2}$ \\
	7 & $0.20\pm0.0$ & $0.42\pm0.1$ & $0.17\pm0.0$ & $0.31\pm0.1$ & $\mathbf{0.16\pm0.0}$ & $\mathbf{0.28\pm0.1}$ & $0.16\pm0.0$ & $0.28\pm0.1$ \\
	8 & $0.29\pm0.1$ & $0.51\pm0.4$ & $\mathbf{0.29\pm0.1}$ & $0.49\pm0.4$ & $0.29\pm0.1$ & $0.48\pm0.3$ & $0.29\pm0.1$ & $\mathbf{0.48\pm0.4}$ \\
	9 & $0.27\pm0.1$ & $0.30\pm0.1$ & $0.27\pm0.1$ & $0.29\pm0.1$ & $0.27\pm0.1$ & $0.28\pm0.1$ & $\mathbf{0.26\pm0.1}$ & $\mathbf{0.28\pm0.1}$ \\
	10 & $0.30\pm0.0$ & $1.04\pm0.1$ & $\mathbf{0.27\pm0.0}$ & $\mathbf{0.89\pm0.1}$ & $0.29\pm0.0$ & $0.89\pm0.1$ & $0.28\pm0.0$ & $0.89\pm0.1$ \LL}%

Regarding the loss function, unmixing results for corresponding pairs of experiments, i.e. experiments with the same architecture and on the same dataset but with different loss functions are comparable. For pairs 1 -- 2, 6 -- 7 and 8 -- 9, and for all weight initialisation methods, both abundances and endmembers errors were slightly smaller when the autoencoder was trained using the SAD loss, compared to the MSE. The lowest average error values are usually achieved for XGU initialisation. Furthermore, for all experiments, this weight initialisation method led to lower mean endmembers error than KHN initialisation. Overall, Glorot initialisation technique seems to be on average better than He methods, when considering abundances / endmembers error values.

For some models with ReLU activation function, due to a large number of dead activations, a problem with the gradient flow during backpropagation steps emerged. This is especially important for bottleneck autoencoder architectures for HU. Presented reinitialisation methods which limit the number of dead activations can alleviate this problem allowing the signal to flow.

\section{Conclusions}

% Hyperparameter optimisation is crucial for effectiveness of a machine learning model. 
Evaluation of models trained with different hyperparameter values assumes that the training processes are stable. In the case of random appearance of undertrained models there is a significant risk of bias in optimised hyperparameter values.

We have explored this phenomenon for the case of HU using autoencoders.  We have observed cases of the vanishing gradient in first AE layers and confirmed that initialisation has a crucial impact on the final AE performance. A weak initialisation leads to high reconstruction or endmembers errors, despite proper values of hyperparameters selected e.g. by RayTune~\cite{liaw2018tune}. The problem was observed under a range of hyperparameter values, datasets, architectures, and initialisation methods. The phenomenon was confirmed by statistical verification, based on a large set of training experiments. Finally, we have presented three AE improvement methods based on a reinitialisation of all or some network weights. Our results are possibly applicable beyond the HU problem, into all domains where similar AE architectures are used. In the future, we would like to look into the impact of reinitialisation techniques on the performance of multilayer perceptrons in classification or regression tasks.

\subsection*{Appendix}
\begin{figure*}[!h]
	\begin{subfigure}[b]{0.235\textwidth}
		\includegraphics[trim={3,3cm, 0,6cm, 3,5cm, 1,2cm},clip,width=\textwidth]{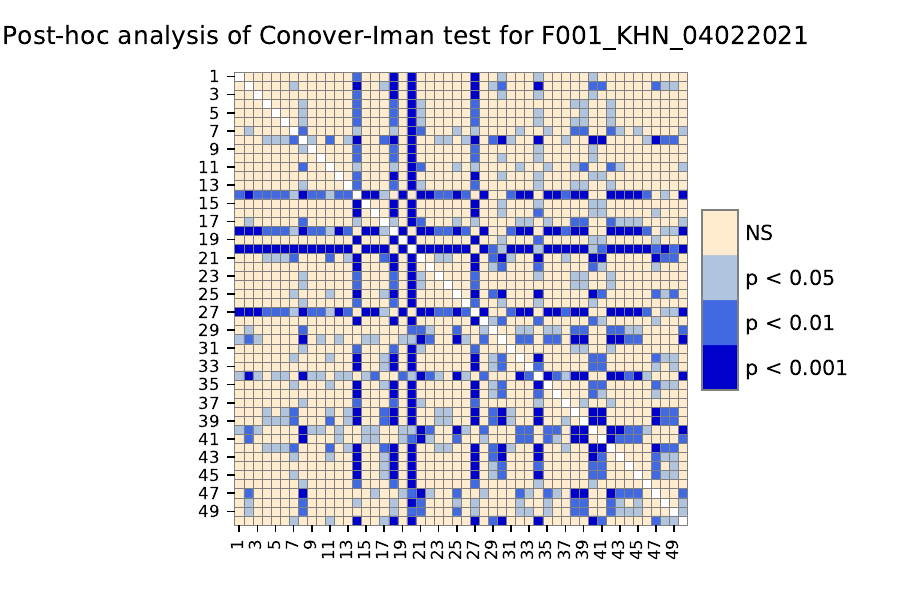}%
		\caption{KHN, exp. 1}
	\end{subfigure}%
	\begin{subfigure}[b]{0.235\textwidth}
		\includegraphics[trim={3,3cm, 0,6cm, 3,5cm, 1,2cm},clip,width=\textwidth]{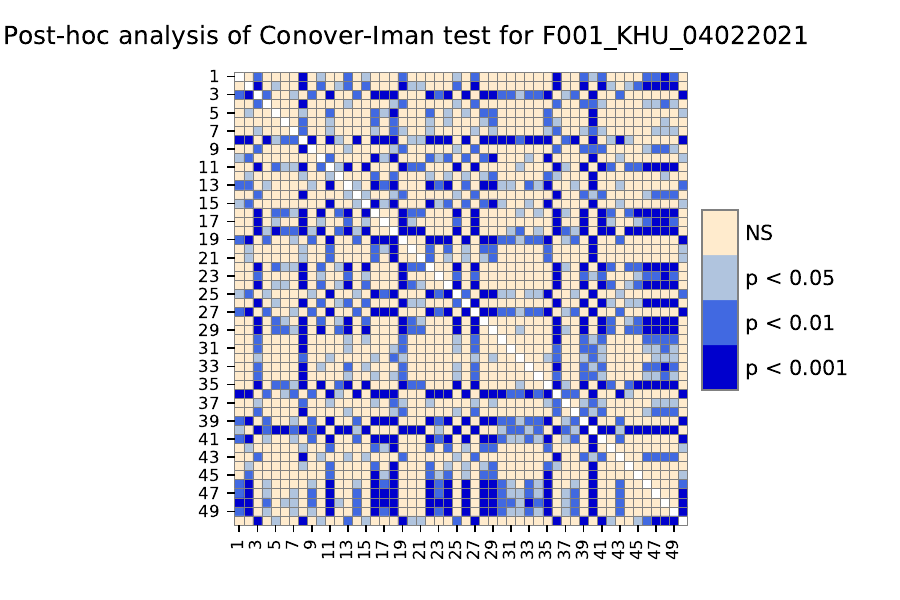}%
		\caption{KHU, exp. 1}
	\end{subfigure}%
	\begin{subfigure}[b]{0.235\textwidth}
		\includegraphics[trim={3,3cm, 0,6cm, 3,5cm, 1,2cm},clip,width=\textwidth]{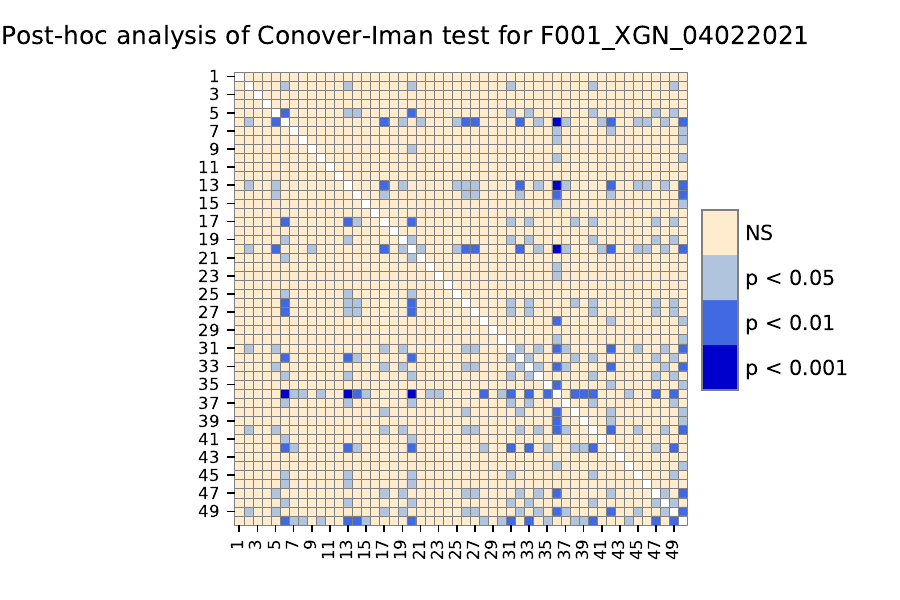}%
		\caption{XGN, exp. 1}
	\end{subfigure}%
	\begin{subfigure}[b]{0.3275\textwidth}
		\includegraphics[trim={3,3cm, 0,6cm, 0cm, 1,2cm},clip,width=\textwidth]{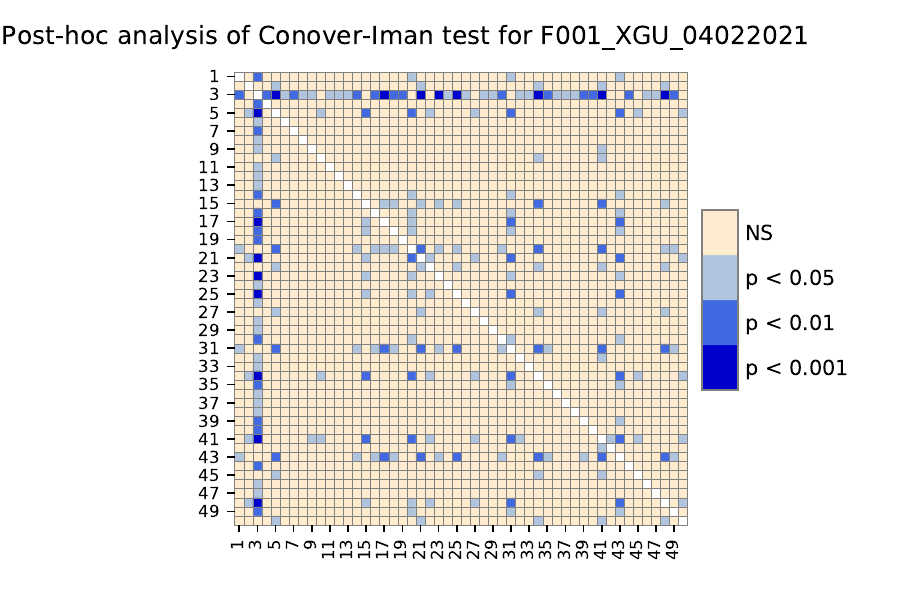}
		\caption{XGU, exp. 1~~~~~~~~~~}
	\end{subfigure}
	\begin{subfigure}[b]{0.235\textwidth}
		\includegraphics[trim={3,3cm, 0,6cm, 3,5cm, 1,2cm},clip,width=\textwidth]{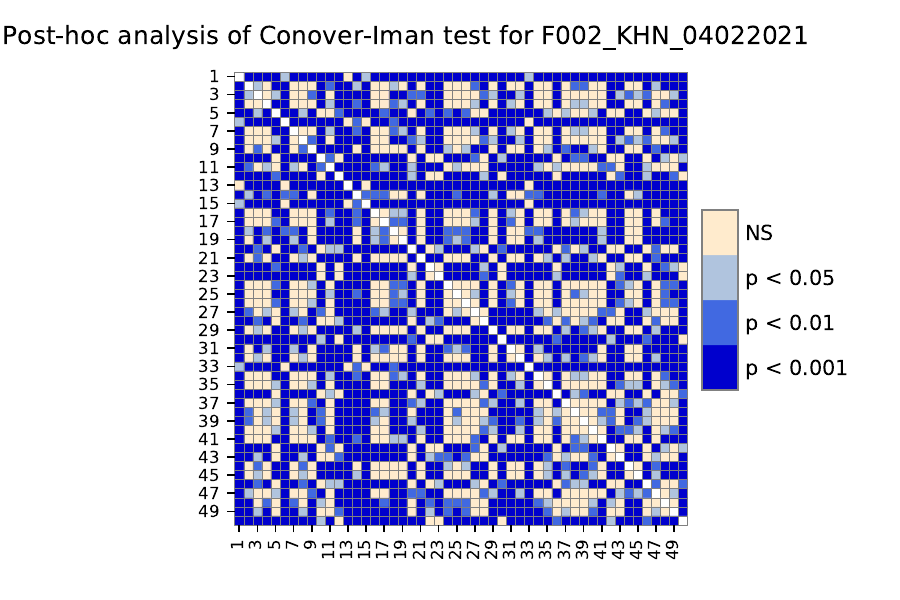}%
		\caption{KHN, exp. 2}
	\end{subfigure}%
	\begin{subfigure}[b]{0.235\textwidth}
		\includegraphics[trim={3,3cm, 0,6cm, 3,5cm, 1,2cm},clip,width=\textwidth]{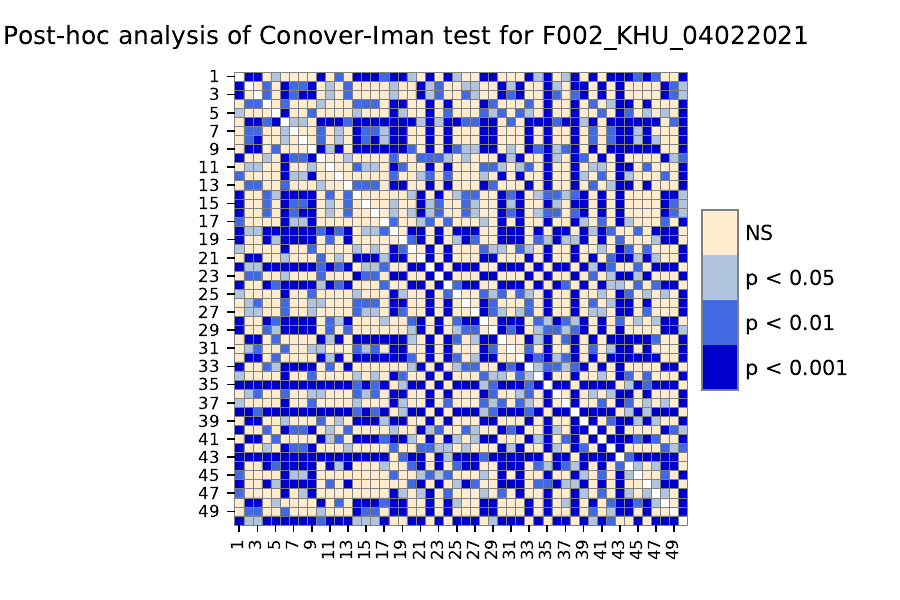}%
		\caption{KHU, exp. 2}
	\end{subfigure}%
	\begin{subfigure}[b]{0.235\textwidth}
		\includegraphics[trim={3,3cm, 0,6cm, 3,5cm, 1,2cm},clip,width=\textwidth]{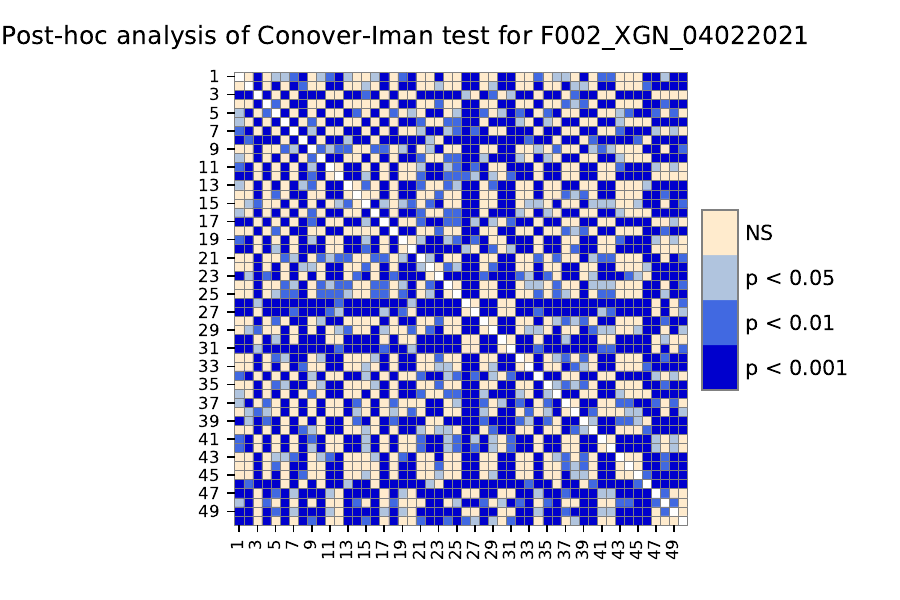}%
		\caption{XGN, exp. 2}
	\end{subfigure}%
	\begin{subfigure}[b]{0.3275\textwidth}
		\includegraphics[trim={3,3cm, 0,6cm, 0cm, 1,2cm},clip,width=\textwidth]{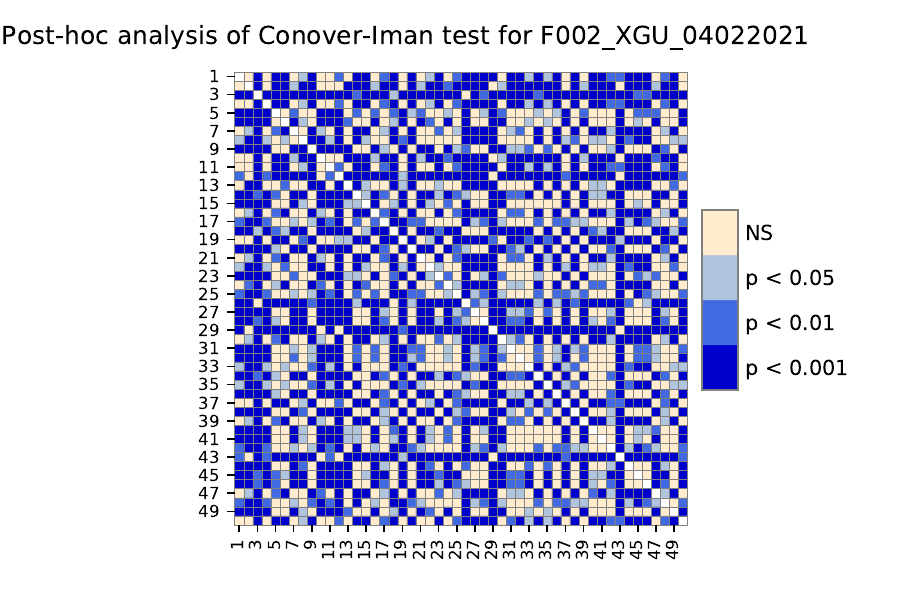}
		\caption{XGU, exp. 2~~~~~~~~~~}
	\end{subfigure}
	\begin{subfigure}[b]{0.235\textwidth}
		\includegraphics[trim={3,3cm, 0,6cm, 3,5cm, 1,2cm},clip,width=\textwidth]{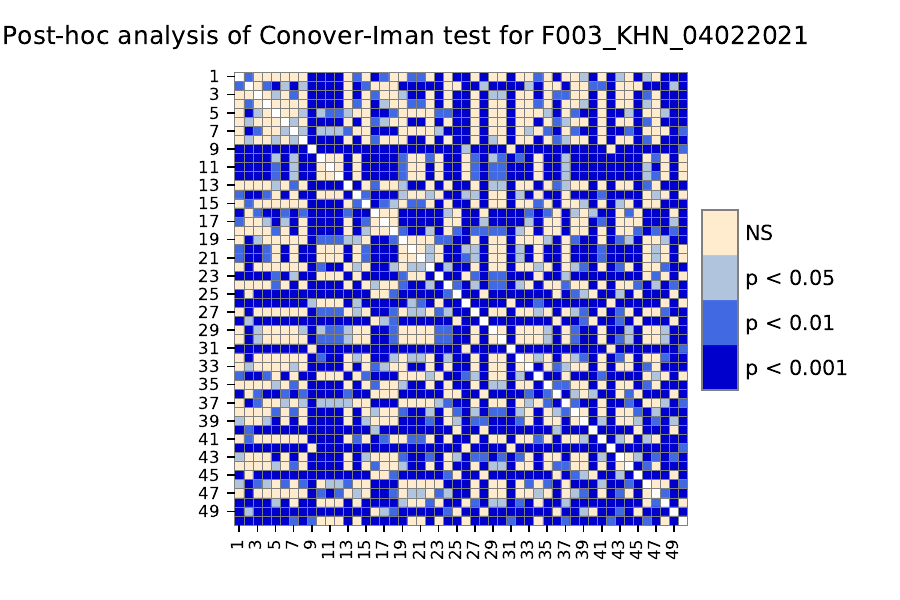}%
		\caption{KHN, exp. 3}
	\end{subfigure}%
	\begin{subfigure}[b]{0.235\textwidth}
		\includegraphics[trim={3,3cm, 0,6cm, 3,5cm, 1,2cm},clip,width=\textwidth]{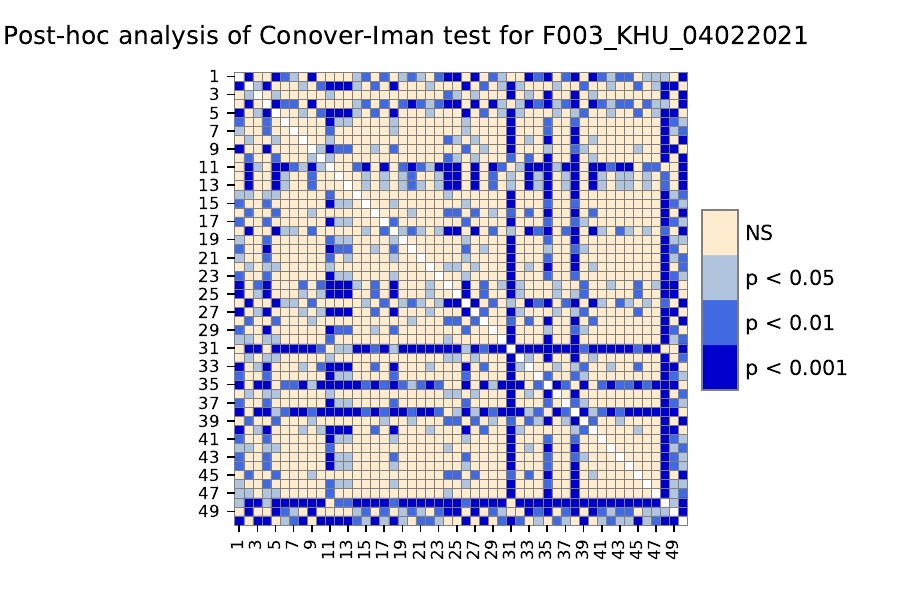}%
		\caption{KHU, exp. 3}
	\end{subfigure}%
	\begin{subfigure}[b]{0.235\textwidth}
		\includegraphics[trim={3,3cm, 0,6cm, 3,5cm, 1,2cm},clip,width=\textwidth]{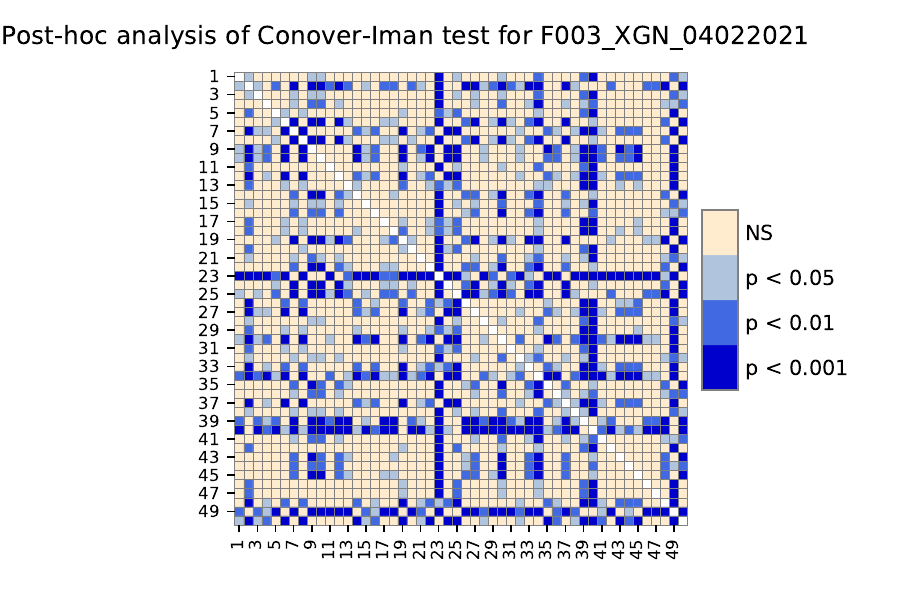}%
		\caption{XGN, exp. 3}
	\end{subfigure}%
	\begin{subfigure}[b]{0.3275\textwidth}
		\includegraphics[trim={3,3cm, 0,6cm, 0cm, 1,2cm},clip,width=\textwidth]{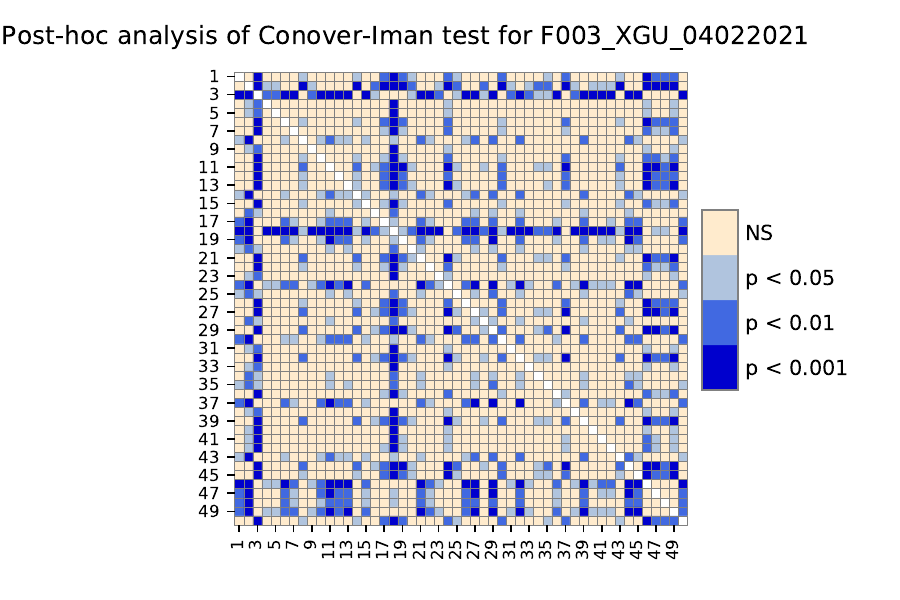}
		\caption{XGU, exp. 3~~~~~~~~~~}
	\end{subfigure}
	\caption{Results of Conover-Iman test for consecutive pairs of models for Experiments 1--3.}
	\label{fig:post_hoc_1-3}
\end{figure*}

\begin{figure*}[!h]
	\begin{subfigure}[b]{0.235\textwidth}
		\includegraphics[trim={3,3cm, 0,6cm, 3,5cm, 1,2cm},clip,width=\textwidth]{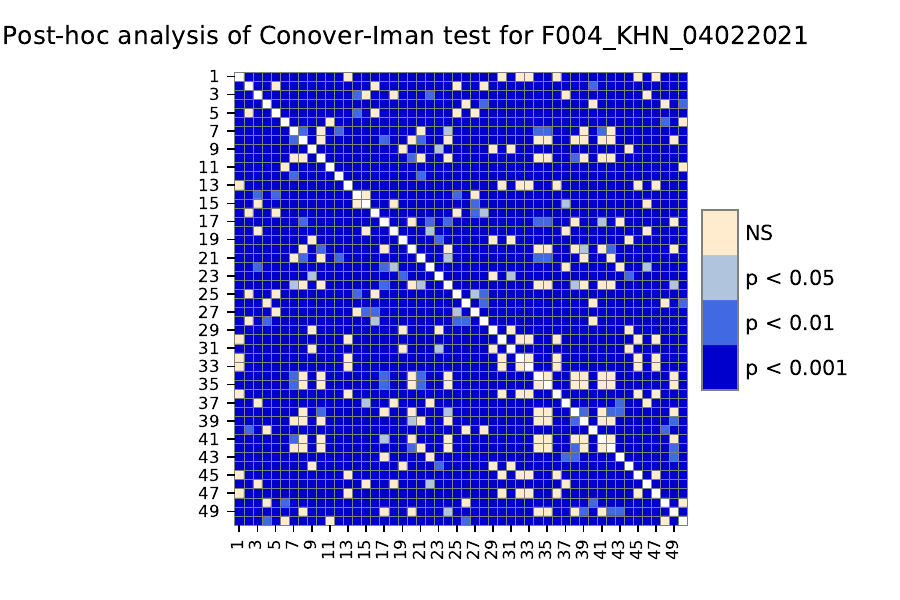}%
		\caption{KHN, exp. 4}
	\end{subfigure}%
	\begin{subfigure}[b]{0.235\textwidth}
		\includegraphics[trim={3,3cm, 0,6cm, 3,5cm, 1,2cm},clip,width=\textwidth]{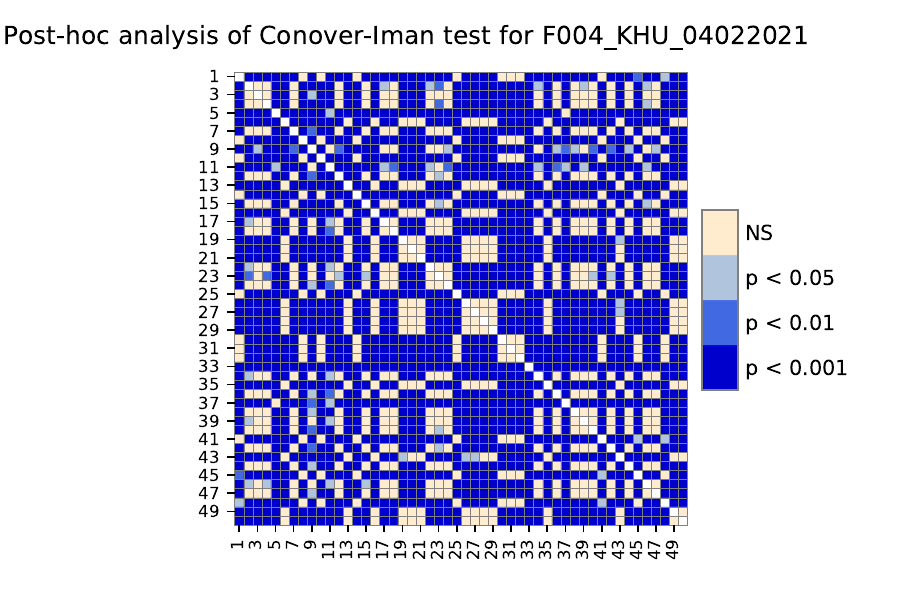}%
		\caption{KHU, exp. 4}
	\end{subfigure}%
	\begin{subfigure}[b]{0.235\textwidth}
		\includegraphics[trim={3,3cm, 0,6cm, 3,5cm, 1,2cm},clip,width=\textwidth]{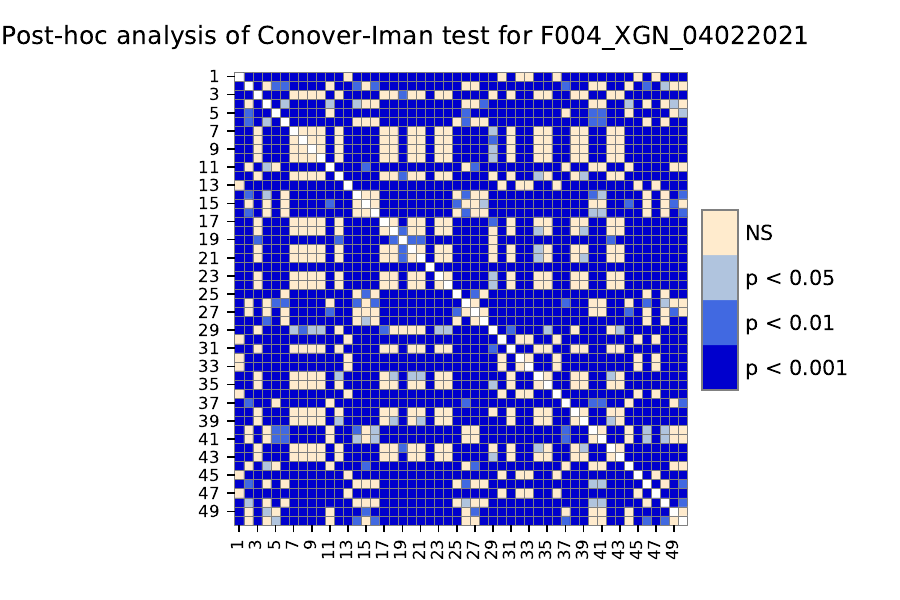}%
		\caption{XGN, exp. 4}
	\end{subfigure}%
	\begin{subfigure}[b]{0.3275\textwidth}
		\includegraphics[trim={3,3cm, 0,6cm, 0cm, 1,2cm},clip,width=\textwidth]{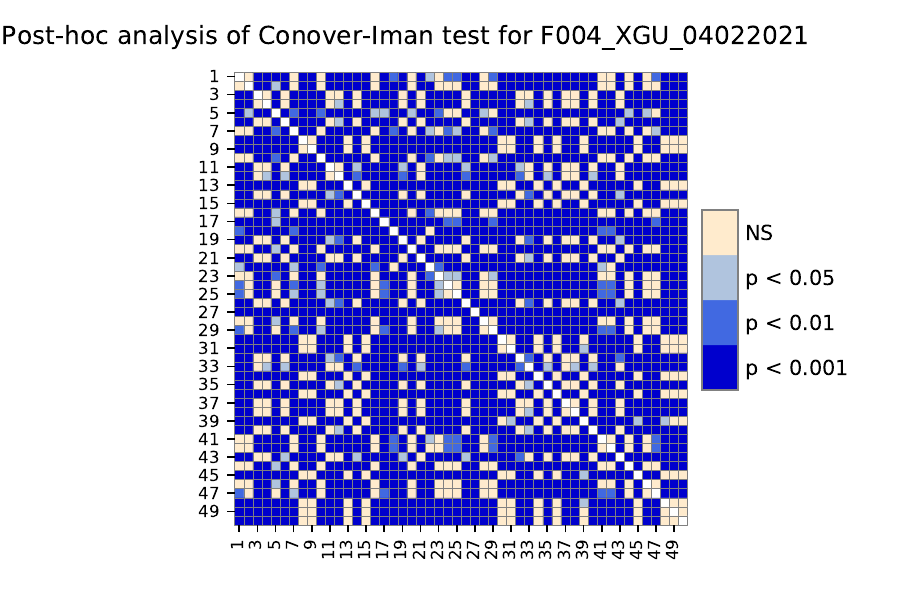}
		\caption{XGU, exp. 4~~~~~~~~~~}
	\end{subfigure}
	\begin{subfigure}[b]{0.235\textwidth}
		\includegraphics[trim={3,3cm, 0,6cm, 3,5cm, 1,2cm},clip,width=\textwidth]{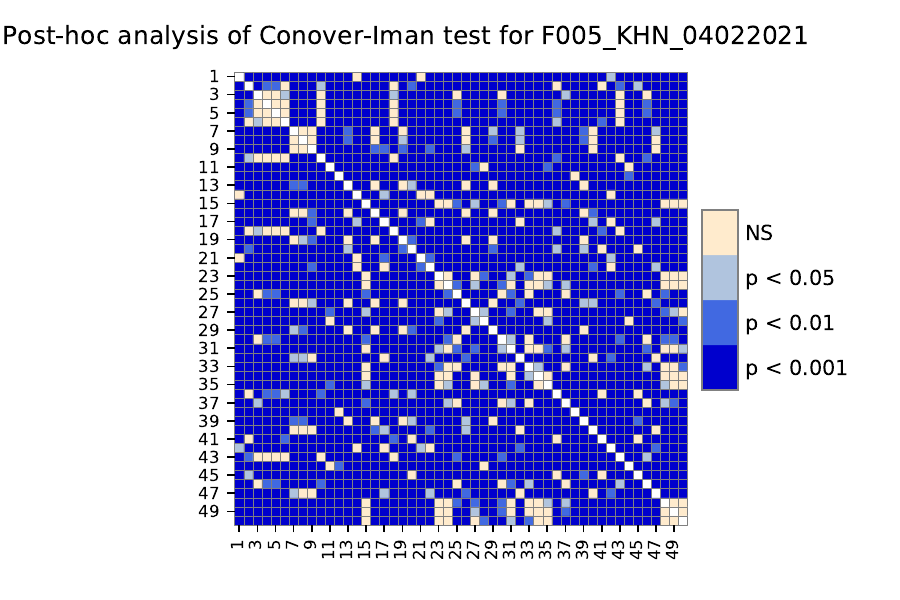}%
		\caption{KHN, exp. 5}
	\end{subfigure}%
	\begin{subfigure}[b]{0.235\textwidth}
		\includegraphics[trim={3,3cm, 0,6cm, 3,5cm, 1,2cm},clip,width=\textwidth]{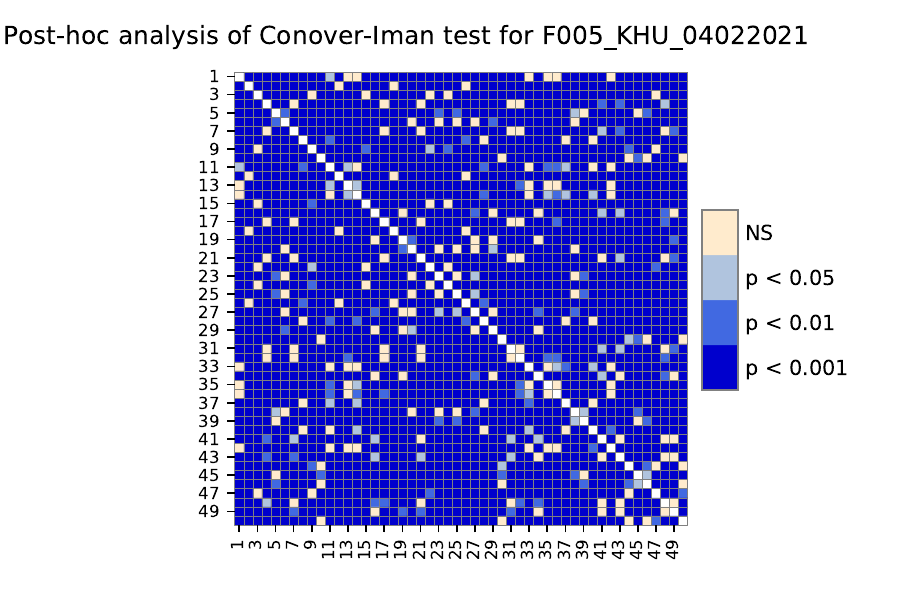}%
		\caption{KHU, exp. 5}
	\end{subfigure}%
	\begin{subfigure}[b]{0.235\textwidth}
		\includegraphics[trim={3,3cm, 0,6cm, 3,5cm, 1,2cm},clip,width=\textwidth]{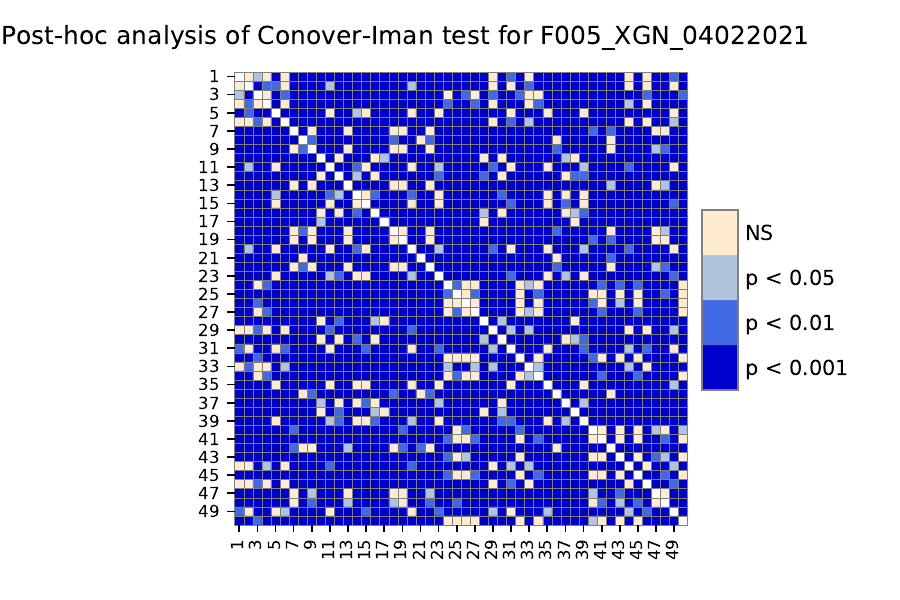}%
		\caption{XGN, exp. 5}
	\end{subfigure}%
	\begin{subfigure}[b]{0.3275\textwidth}
		\includegraphics[trim={3,3cm, 0,6cm, 0cm, 1,2cm},clip,width=\textwidth]{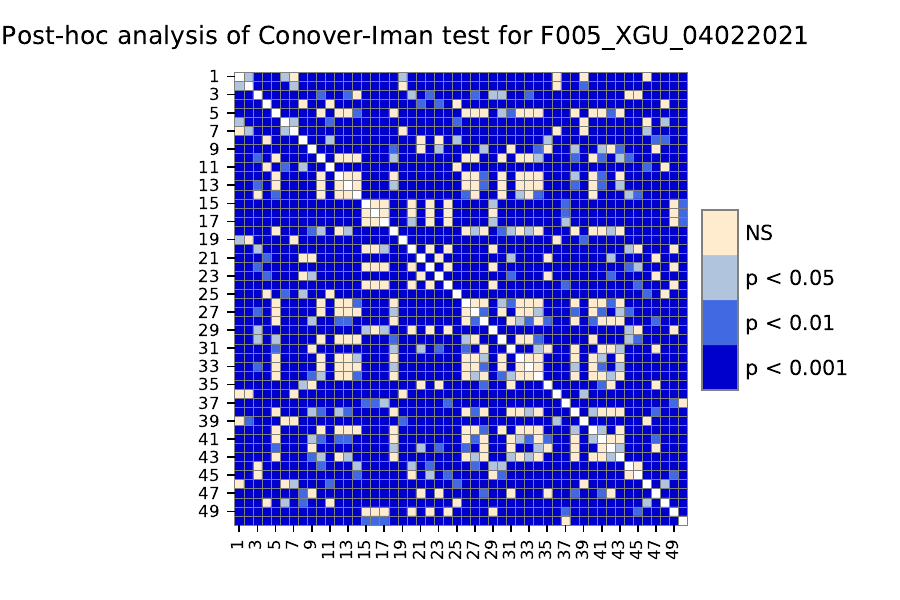}
		\caption{XGU, exp. 5~~~~~~~~~~}
	\end{subfigure}
	\begin{subfigure}[b]{0.235\textwidth}
		\includegraphics[trim={3,3cm, 0,6cm, 3,5cm, 1,2cm},clip,width=\textwidth]{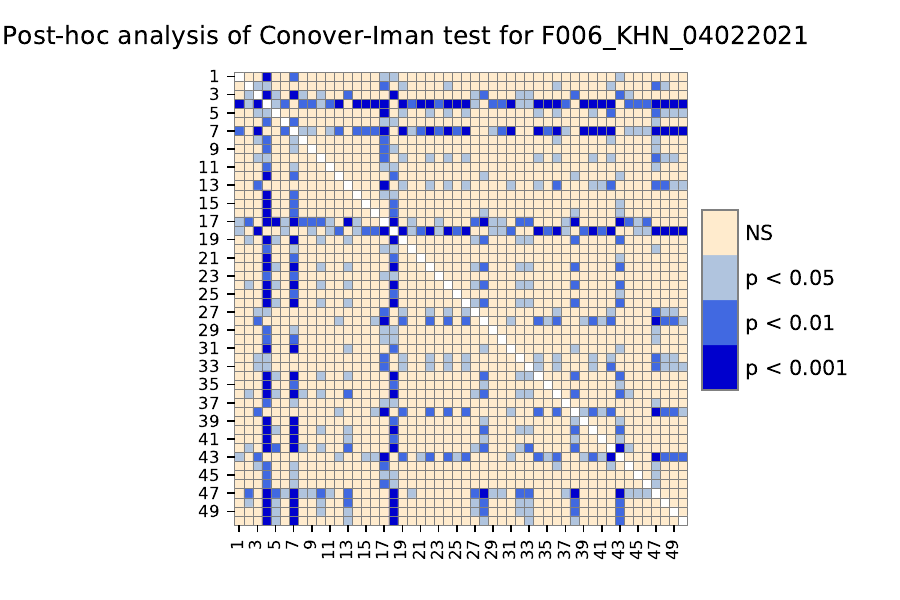}%
		\caption{KHN, exp. 6}
	\end{subfigure}%
	\begin{subfigure}[b]{0.235\textwidth}
		\includegraphics[trim={3,3cm, 0,6cm, 3,5cm, 1,2cm},clip,width=\textwidth]{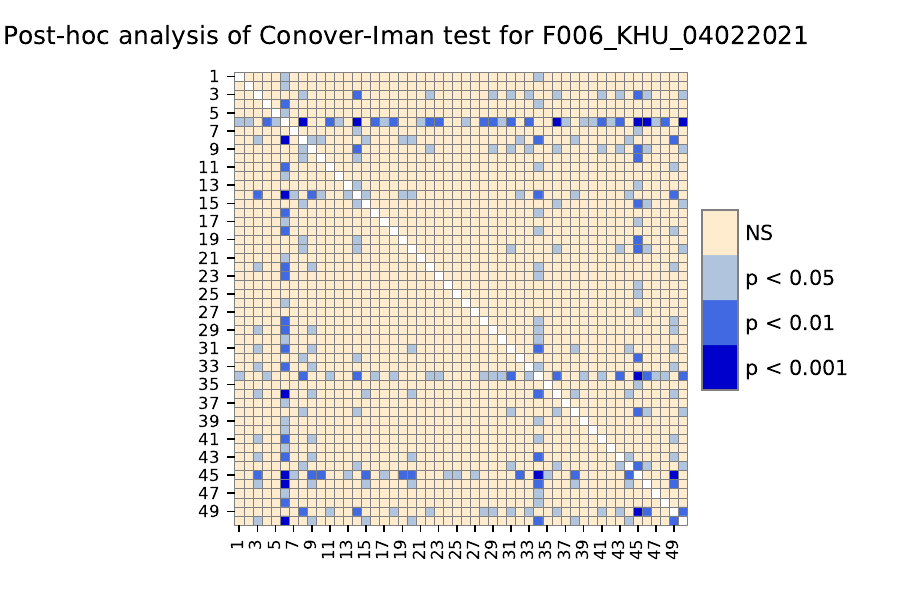}%
		\caption{KHU, exp. 6}
	\end{subfigure}%
	\begin{subfigure}[b]{0.235\textwidth}
		\includegraphics[trim={3,3cm, 0,6cm, 3,5cm, 1,2cm},clip,width=\textwidth]{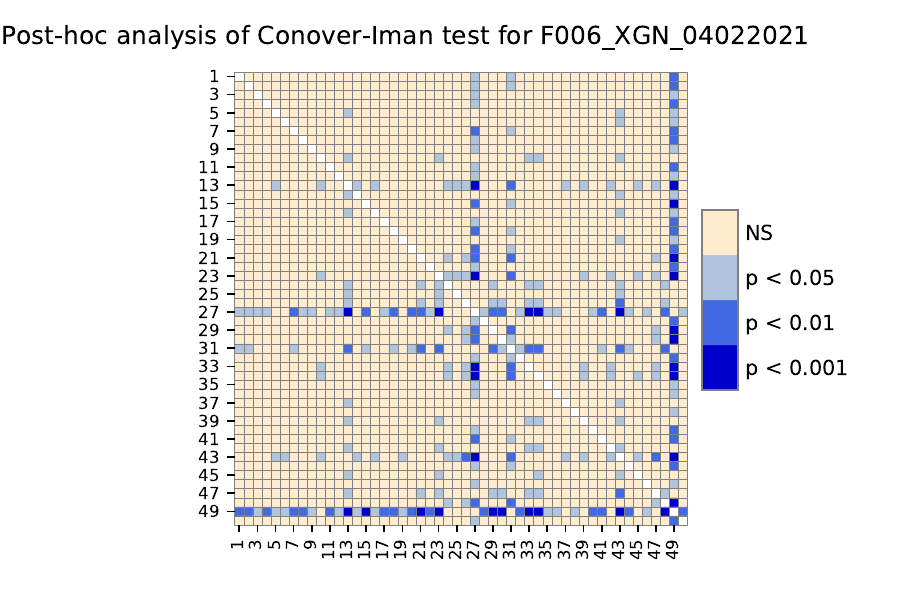}%
		\caption{XGN, exp. 6}
	\end{subfigure}%
	\begin{subfigure}[b]{0.3275\textwidth}
		\includegraphics[trim={3,3cm, 0,6cm, 0cm, 1,2cm},clip,width=\textwidth]{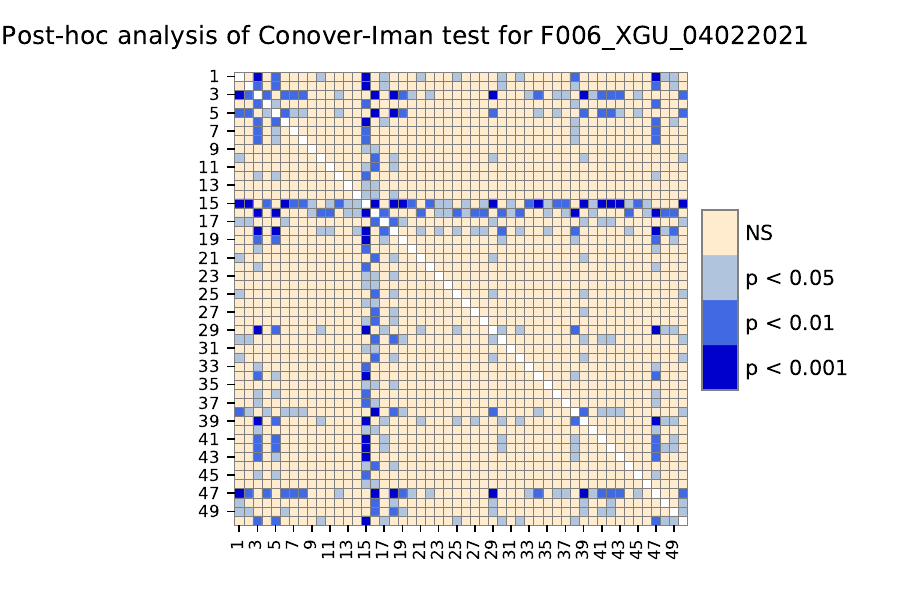}
		\caption{XGU, exp. 6~~~~~~~~~~}
	\end{subfigure}
	\begin{subfigure}[b]{0.235\textwidth}
		\includegraphics[trim={3,3cm, 0,6cm, 3,5cm, 1,2cm},clip,width=\textwidth]{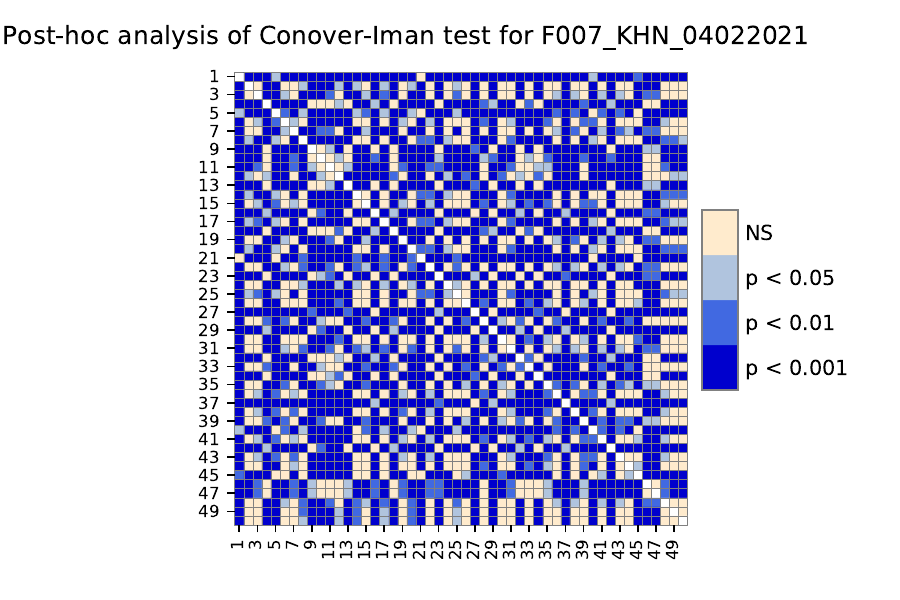}%
		\caption{KHN, exp. 7}
	\end{subfigure}%
	\begin{subfigure}[b]{0.235\textwidth}
		\includegraphics[trim={3,3cm, 0,6cm, 3,5cm, 1,2cm},clip,width=\textwidth]{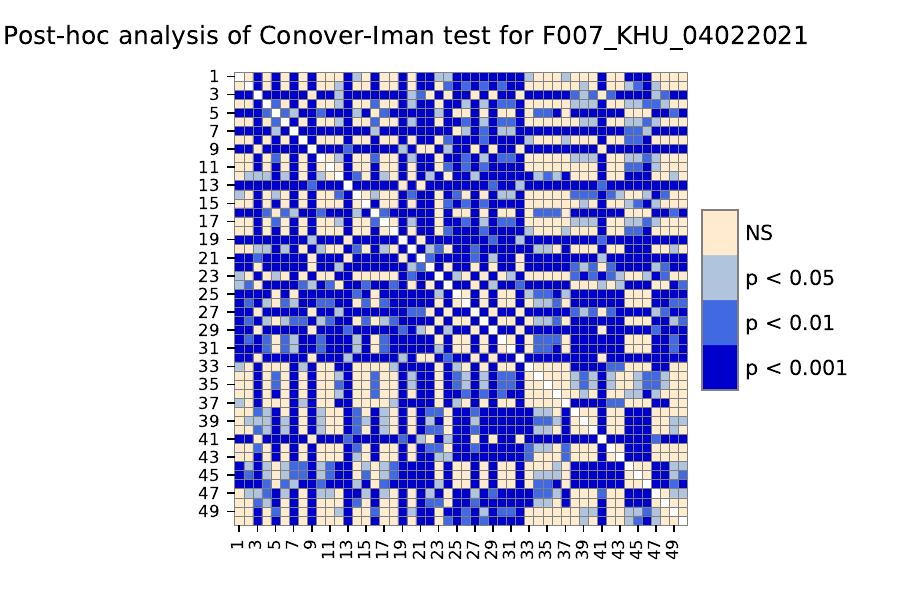}%
		\caption{KHU, exp. 7}
	\end{subfigure}%
	\begin{subfigure}[b]{0.235\textwidth}
		\includegraphics[trim={3,3cm, 0,6cm, 3,5cm, 1,2cm},clip,width=\textwidth]{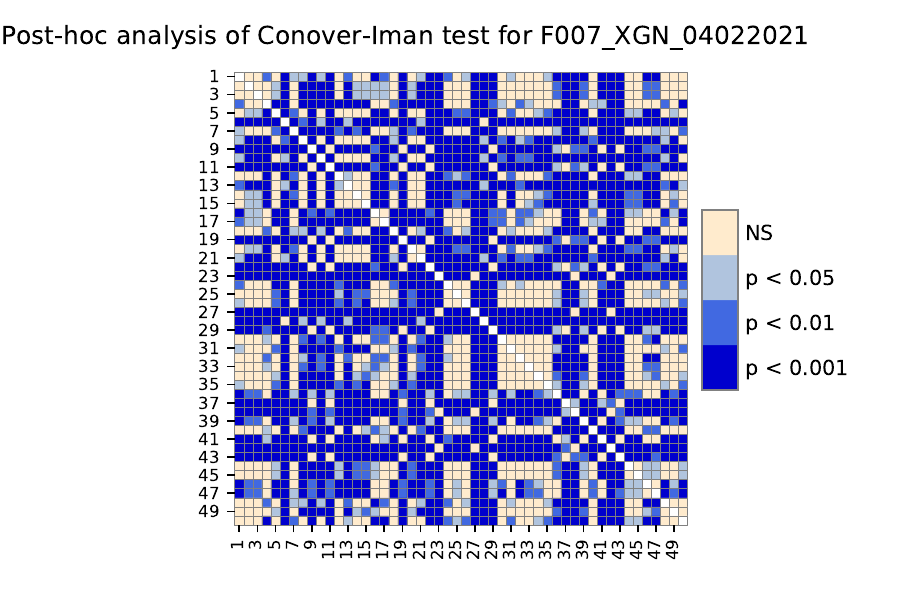}%
		\caption{XGN, exp. 7}
	\end{subfigure}%
	\begin{subfigure}[b]{0.3275\textwidth}
		\includegraphics[trim={3,3cm, 0,6cm, 0cm, 1,2cm},clip,width=\textwidth]{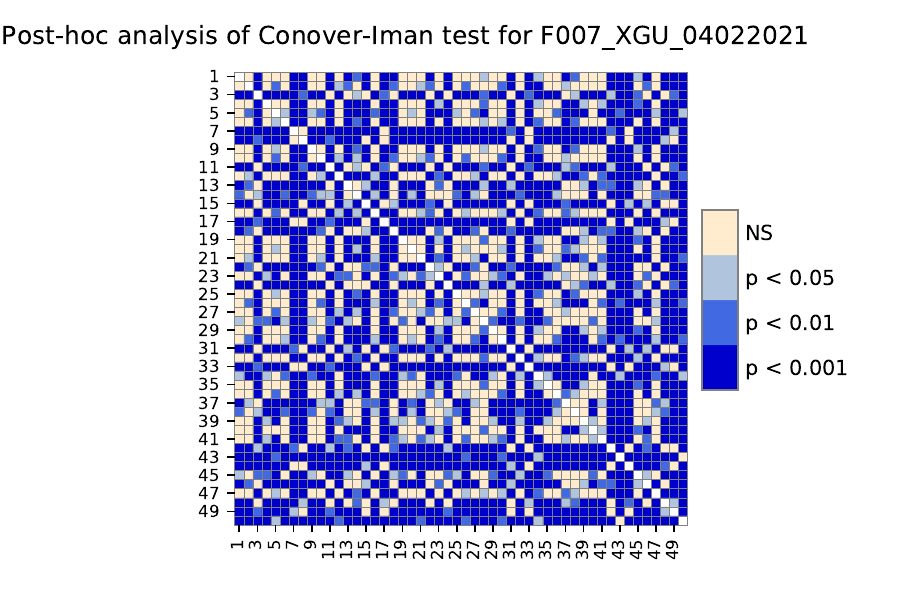}
		\caption{XGU, exp. 7~~~~~~~~~~}
	\end{subfigure}
	\begin{subfigure}[b]{0.235\textwidth}
		\includegraphics[trim={3,3cm, 0,6cm, 3,5cm, 1,2cm},clip,width=\textwidth]{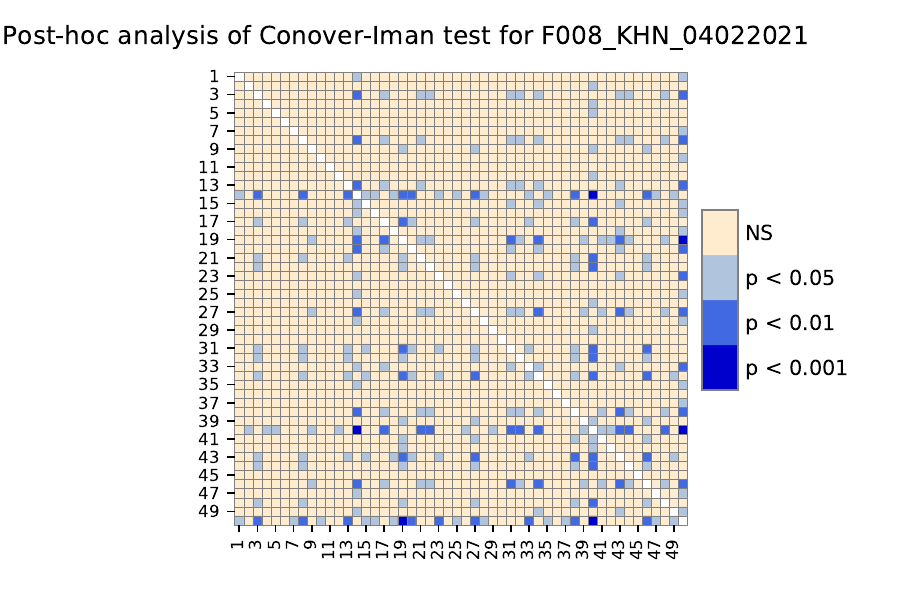}%
		\caption{KHN, exp. 8}
	\end{subfigure}%
	\begin{subfigure}[b]{0.235\textwidth}
		\includegraphics[trim={3,3cm, 0,6cm, 3,5cm, 1,2cm},clip,width=\textwidth]{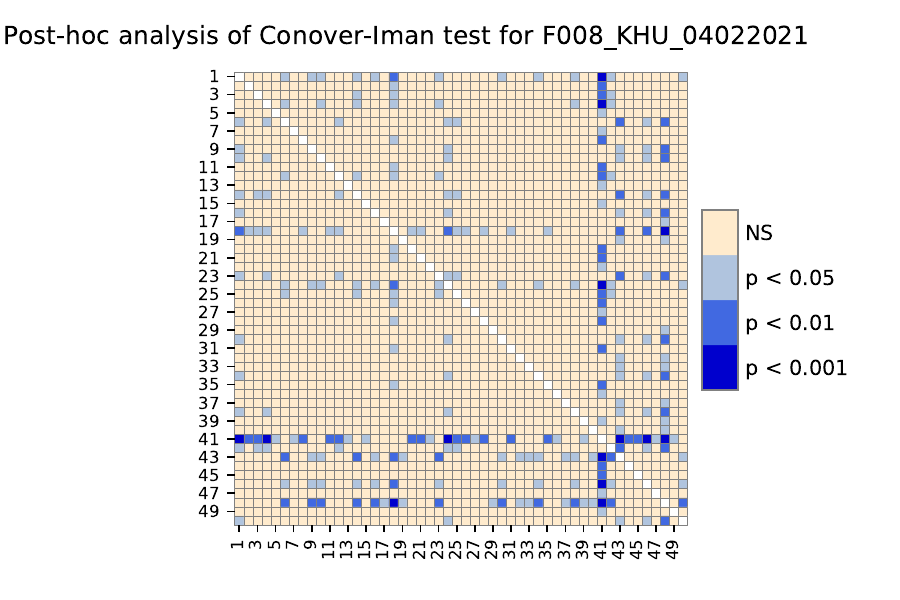}%
		\caption{KHU, exp. 8}
	\end{subfigure}%
	\begin{subfigure}[b]{0.235\textwidth}
		\includegraphics[trim={3,3cm, 0,6cm, 3,5cm, 1,2cm},clip,width=\textwidth]{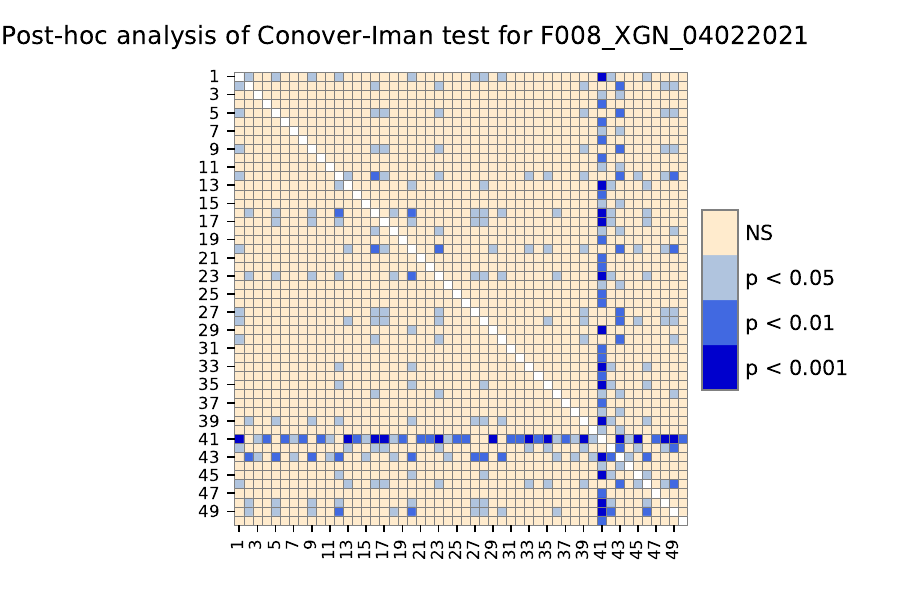}%
		\caption{XGN, exp. 8}
	\end{subfigure}%
	\begin{subfigure}[b]{0.3275\textwidth}
		\includegraphics[trim={3,3cm, 0,6cm, 0cm, 1,2cm},clip,width=\textwidth]{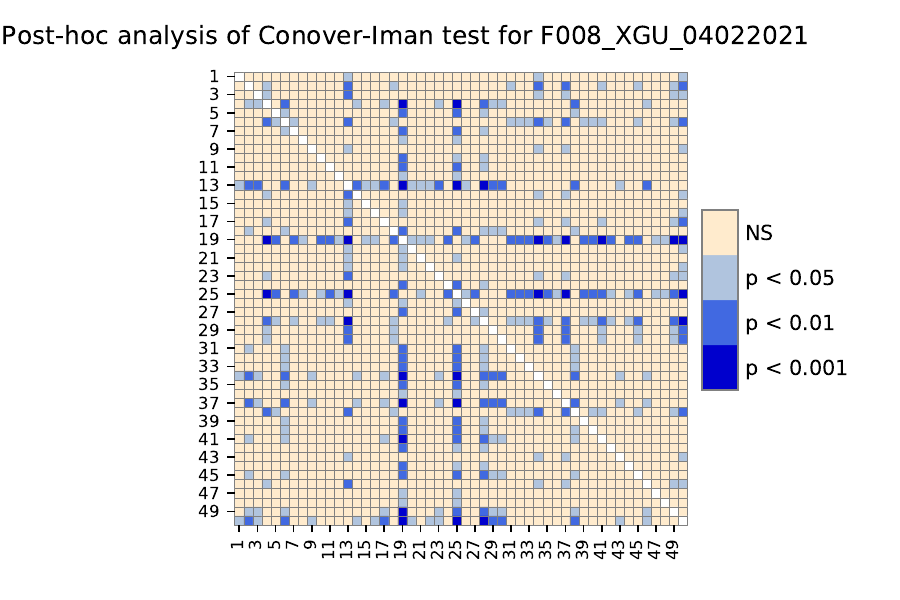}
		\caption{XGU, exp. 8~~~~~~~~~~}
	\end{subfigure}
	\caption{Results of Conover-Iman test for consecutive pairs of models for Experiments 4--8.}
	\label{fig:post_hoc_4-8}
\end{figure*}

\begin{figure*}[!h]
	\begin{subfigure}[b]{0.235\textwidth}
		\includegraphics[trim={3,3cm, 0,6cm, 3,5cm, 1,2cm},clip,width=\textwidth]{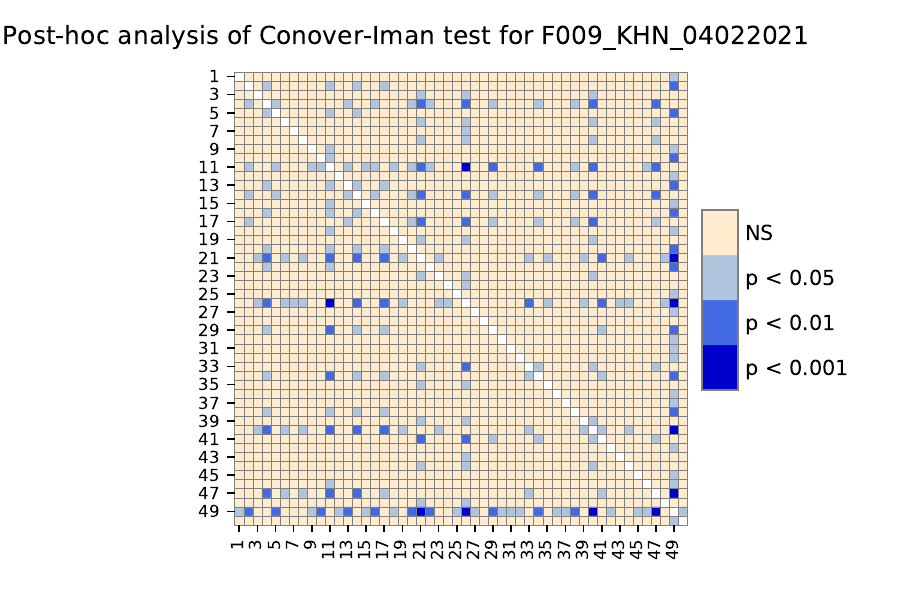}%
		\caption{KHN, exp. 9}
	\end{subfigure}%
	\begin{subfigure}[b]{0.235\textwidth}
		\includegraphics[trim={3,3cm, 0,6cm, 3,5cm, 1,2cm},clip,width=\textwidth]{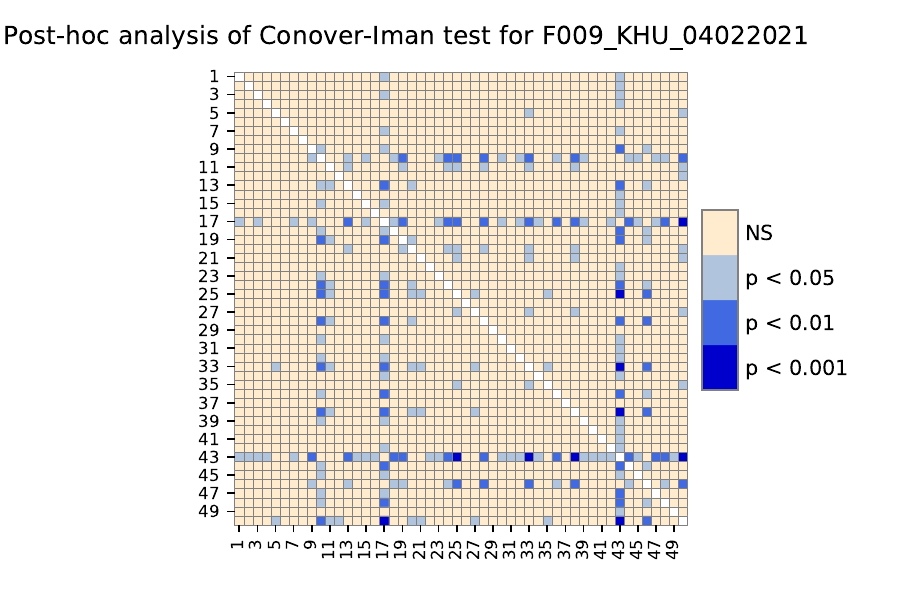}%
		\caption{KHU, exp. 9}
	\end{subfigure}%
	\begin{subfigure}[b]{0.235\textwidth}
		\includegraphics[trim={3,3cm, 0,6cm, 3,5cm, 1,2cm},clip,width=\textwidth]{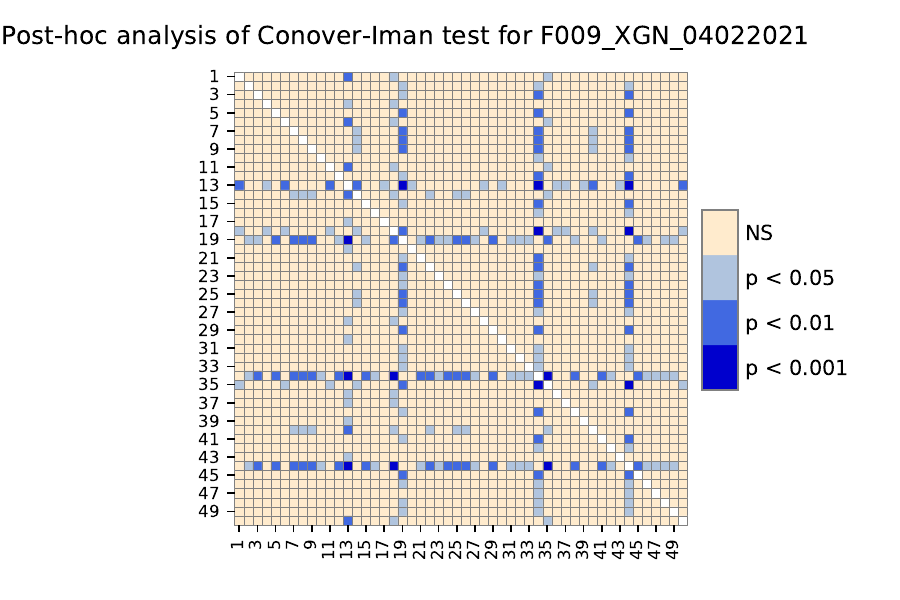}%
		\caption{XGN, exp. 9}
	\end{subfigure}%
	\begin{subfigure}[b]{0.3275\textwidth}
		\includegraphics[trim={3,3cm, 0,6cm, 0cm, 1,2cm},clip,width=\textwidth]{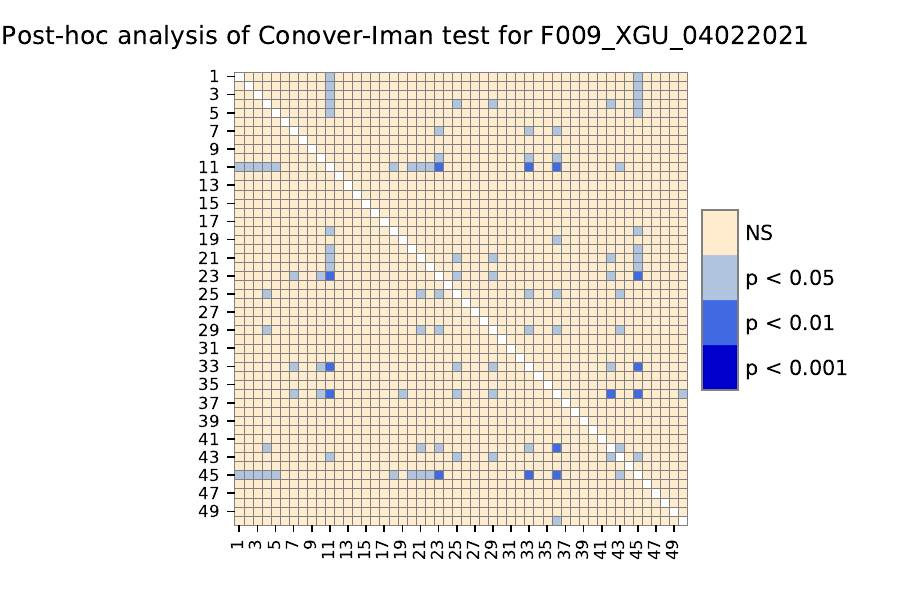}
		\caption{XGU, exp. 9~~~~~~~~~~}
	\end{subfigure}
	\begin{subfigure}[b]{0.235\textwidth}
		\includegraphics[trim={3,3cm, 0,6cm, 3,5cm, 1,2cm},clip,width=\textwidth]{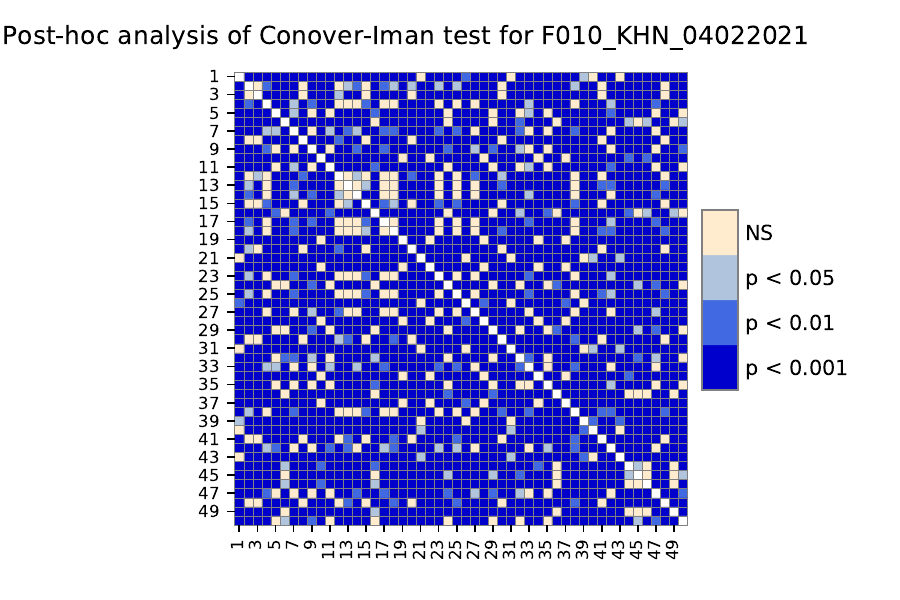}%
		\caption{KHN, exp. 10}
	\end{subfigure}%
	\begin{subfigure}[b]{0.235\textwidth}
		\includegraphics[trim={3,3cm, 0,6cm, 3,5cm, 1,2cm},clip,width=\textwidth]{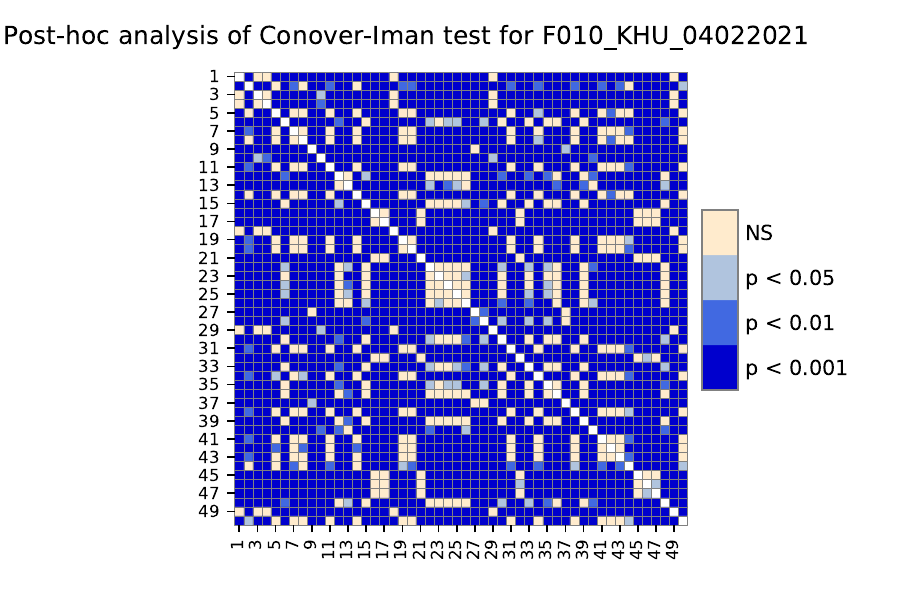}%
		\caption{KHU, exp. 10}
	\end{subfigure}%
	\begin{subfigure}[b]{0.235\textwidth}
		\includegraphics[trim={3,3cm, 0,6cm, 3,5cm, 1,2cm},clip,width=\textwidth]{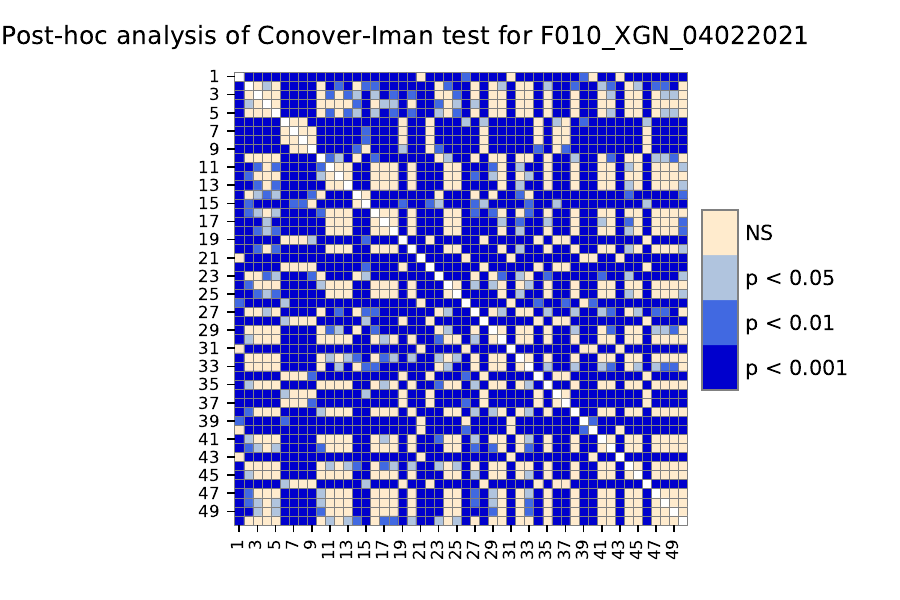}%
		\caption{XGN, exp. 10}
	\end{subfigure}%
	\begin{subfigure}[b]{0.3275\textwidth}
		\includegraphics[trim={3,3cm, 0,6cm, 0cm, 1,2cm},clip,width=\textwidth]{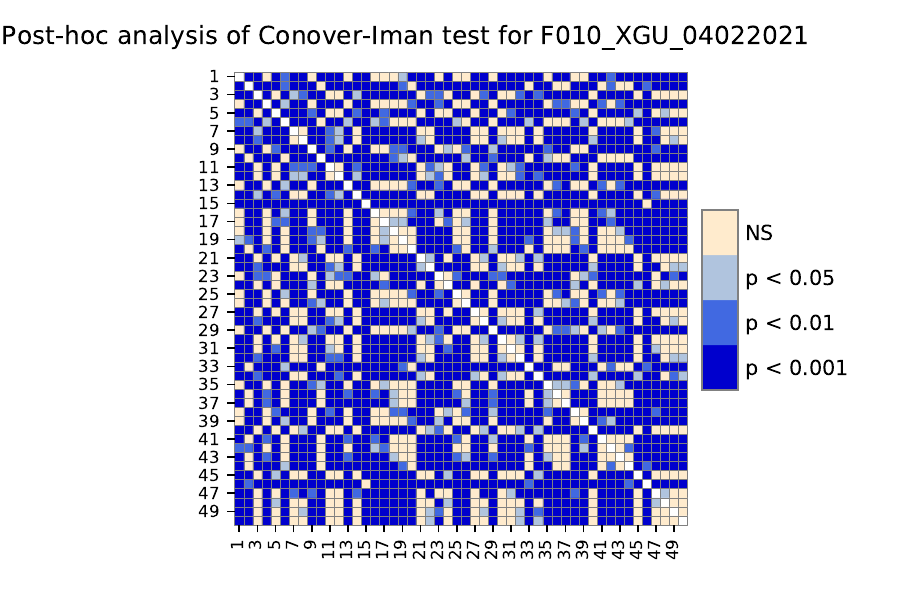}
		\caption{XGU, exp. 10~~~~~~~~~~}
	\end{subfigure}
	\caption{Results of Conover-Iman test for consecutive pairs of models for Experiments 9--10.}
	\label{fig:post_hoc_9-10}
\end{figure*}

Figures~\ref{fig:post_hoc_1-3}--\ref{fig:post_hoc_9-10} depict results of post-hoc Conover-Iman test for pairs of models for all experiments conducted. The test has been performed for RMSE values. A heat-map cell located at position $(i, j)$, where $i, j \in \lbrace 1, 2, ..., 50 \rbrace$, indicates the statistical significance of a difference between the $i$--th and the $j$--th model of a given experiment scenario, i.e. $50$ runs of the trained models. The darker the cell color is, the smaller the p-value. \emph{NS} means that the difference between models is not statistically significant, in terms of RMSE values. KHN and KHU denote He initialisation with normal and uniform distribution, respectively, while XGN and XGU represent Glorot initialisation method, also with normal or uniform distribution.

Generally, for a given scenario, test results for all weights initialisation methods are similar. The situation is different in the case of Experiments 1 and 3. For Experiment 1, Glorot initialisations were more stable than He approaches while for Experiment 3, KHN led to the largest number of outlying models. It is also possible to conclude that the highest rate of statistically significant differences between models was achieved for Experiments 4, 5 and 10.

\subsubsection*{Acknowledgements} K.K. acknowledges funding from the European Union through the European Social Fund
(grant POWR.03.02.00-00-I029). B.G. acknowledges funding from the budget funds for science in the years 2018-2022, as a scientific project ''Application of transfer learning methods in the problem of hyperspectral images classification using convolutional neural networks'' under the ''Diamond Grant'' program, no. DI2017 013847.

	{\small
		\bibliographystyle{ieeetr}
		\bibliography{bibliography}
	}
\end{document}